\documentclass[11pt]{article}
\usepackage{titling}
\usepackage[utf8]{inputenc}
\usepackage[english]{babel}
\usepackage{graphicx}
\usepackage{subcaption}
\usepackage{amsmath, amsfonts, amssymb, amsthm, bm, graphicx, mathtools, enumerate,multirow}
\usepackage[affil-it]{authblk}
\usepackage[round]{natbib}
\usepackage{bbm}
\usepackage{booktabs}
\usepackage{tikz}
\usepackage{relsize}
\usepackage{xr}

\usepackage[CJKbookmarks=true,
bookmarksnumbered=true,
bookmarksopen=true,
colorlinks=true,
citecolor=blue,
linkcolor=blue,
anchorcolor=blue,
urlcolor=blue]{hyperref}
\usepackage[letterpaper, left=1.2truein, right=1.2truein, top = 1.2truein,bottom = 1.2truein]{geometry}
\usepackage[ruled, vlined, lined, commentsnumbered,linesnumbered]{algorithm2e}
\usepackage{prettyref,soul}
\usepackage{enumitem}
\setlist{nolistsep}

\usepackage{xr}
\DeclareMathOperator*{\argmin}{argmin}
\newtheorem{lemma}{Lemma}
\newtheorem{proposition}{Proposition}
\newtheorem{thm}{Theorem}
\newtheorem{definition}{Definition}

\newtheorem{assumption}{Assumption}
\theoremstyle{definition}
\newtheorem{remark}{Remark}
\newtheorem{example}{Example}
\usepackage{float}
\floatstyle{plaintop}
\restylefloat{table}

\def\E{\mathbb{E}} 

\newcommand*{\Scale}[2][4]{\scalebox{#1}{$#2$}}%

\newcommand{\tp}{\intercal}

\newcommand{\bigO}{\ensuremath{\mathop{}\mathopen{}\mathcal{O}\mathopen{}}}

\newcommand{\bigOp}{\bigO_\mathrm{p}}

\newcommand{\ind}{\mbox{$\perp\!\!\!\perp$}}

\newcommand{\Hscr}{{\mathcal{H}}}

\newcommand{\kernel}{K}

\newcommand{\truew}{w^*}
\newcommand{\estw}{\hat w}
\newcommand{\testw}{\tilde w}

\newcommand{\nw}{\texttt{NW}}
\newcommand{\fcbps}{\texttt{FCBPS}}
\newcommand{\npfcbps}{\texttt{NPFCBPS}}
\newcommand{\proposed}{\texttt{CFB}}
\newcommand{\reg}{\texttt{REG}}

\newcommand{\modify}[1]{#1}

\newcounter{ucnt}
\newcommand{\newu}{
	\refstepcounter{ucnt}
	\ensuremath{C_{\theucnt}}
}
\newcommand{\oldu}[1]{\ensuremath{C_{\ref*{#1}}}}

\newcounter{lcnt}
\newcommand{\newl}{
	\refstepcounter{lcnt}
	\ensuremath{c_{\thelcnt}}
}
\newcommand{\oldl}[1]{\ensuremath{c_{\ref*{#1}}}}

\makeatletter

\newcommand{\ltxlabel}[1]{\ltx@label{#1}}

\makeatother

\makeatletter

\newcommand*{\addFileDependency}[1]{
\typeout{(#1)}
\@addtofilelist{#1}
\IfFileExists{#1}{}{\typeout{No file #1.}}
}\makeatother

\title{Flexible Functional Treatment Effect Estimation}

 \author[1]{Jiayi Wang}
 \author[2]{Raymond K. W. Wong}
 \author[3]{Xiaoke Zhang}
 \author[4]{Kwun Chuen Gary Chan}
 
 \affil[1]{Department of Mathematical Sciences, University of Texas at Dallas}
 \affil[2]{Department of Statistics, Texas A\&M University}
 \affil[3]{Department of Statistics, George Washington University}
 \affil[4]{Department of Biostatistics, University of Washington}
\date{}

\begin{document}
\maketitle

\begin{abstract}
    We study treatment effect estimation with functional treatments where the average potential outcome functional is a function of functions, in contrast to continuous treatment effect estimation where the target is a function of real numbers.  By considering a flexible scalar-on-function marginal structural model, a weight-modified kernel ridge regression (WMKRR) is adopted for estimation.  The weights are constructed by directly minimizing the uniform balancing error resulting from a decomposition of the WMKRR estimator, instead of being estimated under a particular treatment selection model.  Despite the complex structure of the uniform balancing error derived under WMKRR, finite-dimensional convex algorithms can be applied to efficiently solve for the proposed weights thanks to a representer theorem. 
    The optimal convergence rate is shown to be attainable by the proposed WMKRR estimator without any smoothness assumption on the true weight function. Corresponding empirical performance is 
    demonstrated by a simulation study and a real data application.
    \end{abstract}
    

    \section{Introduction}
    \label{sec:intro}
    It is well known that for observational studies where a treatment is not randomly assigned, the estimation of average potential outcomes or contrasts (such as average treatment effects) is challenging due to
    possible confoundedness.  This work focuses on estimating the 
    treatment effect with a \textit{functional} treatment, in contrast to the vast majority of existing work that focuses on binary and continuous treatments.
    For motivational purposes, we present a few examples as follows.
    To investigate the causal effect of temperature patterns in a year on crop yields in the following harvest season, one could use daily maximum and daily minimum temperature trajectories as functional treatments \citep{wong2019partially}.  To assess human visceral adipose tissue,
    \cite{zhang2021covariate} used body shape as a functional treatment
    and studied its causal effect on the tissue.
    In addition, biomedical researchers may be interested in the causal effect of the activity profile on certain health indicators, 
    such as body mass index and waist circumference, which are potential indicators of obesity level \citep[][]{neovius2005bmi}.  The activity pattern of an individual can be recorded by a tracker during a certain time period, e.g., as in the Physical Activity Monitor data from the National Health and Nutrition Examination Survey (NHANES) in 2005-2006. One could transform a trajectory of activity intensity into a distribution of the intensity values (respresented by kernel mean embedding \citep{Muandet-Fukumizu-Sriperumbudur17}) and take it as a functional treatment. See Example \ref{exp:kmm} and Section \ref{sec:real} for more details.

    There exists extensive literature on the estimation of average treatment effect (ATE) for binary treatments, which is well summarized by several review papers \citep[e.g.,][]{imbens2004nonparametric,stuart2010matching,ding2018causal,yao2021survey}.
    Extensions to multi-level categorical treatments
    \citep[e.g.,][]{yang2016propensity,lopez2017estimation,li2019propensity}
    and continuous treatments
    \citep[e.g.,][]{robins2000marginal,hirano2004propensity, imai2004causal, imai2014covariate,galvao2015uniformly,zhu2015boosting, kennedy2017non, fong2018covariate,li2020continuous,bahadori2022end} 
    are also abundant.
    Although there is practical interest in the causal effect of functional variables, existing methods that can be applied directly to functional treatments are scarce.
    To the best of our knowledge, only three related works \citep{zhao2018functional,zhang2021covariate, tan2022causal} are devoted to estimating the causal effects of functional treatments using observational data, but each has limitations that we seek to address.
    We will contrast our proposed method with these works in detail below.
    We note that functional variables are sometimes treated as confounders \citep[e.g.,][]{ mckeague2014estimation, laber2018functional, ciarleglio2018constructing,miao2022average}
    and outcomes \citep{zhao2018functional,lin2021causal}. %
    Such settings are fundamentally different from ours, where the functional variables are treatments, accompanied with vector-valued covariates and a real-valued outcome.

    With an unconfoundedness assumption, two modeling strategies are commonly adopted to estimate causal effects.
    One is the outcome regression approach, which first estimates an outcome regression model by treating both the treatments and confounders as predictors, and then averages the regression prediction on a fixed treatment value over the observed covariate distribution. 
    The functional linear model, the most common scalar-on-function model, is often employed for this purpose
    \citep{zhao2018functional}. 
    However, due to the limited flexibility in characterizing the outcome given the functional treatment and multivariate covariates using the functional linear model, inconsistent estimation can arise from model misspecifications.  Although some complex scalar-on-function regression models are available, model misspecification still pose a major risk for causal effect estimation.
    The other approach is based on estimating weights that directly address the selection bias of functional treatments.  Compared to the regression approach, weighting methods try to mimic randomized experiments that theoretically balance on all pre-treatment-assignment variables, and  do not involve the direct use of outcome data in constructing weights. As a result, they help to preserve the objectivity of the analysis and  avoid data snooping \citep{rubin2007design,rosenbaum2010design}. We adopt the weighting approach in this paper.

    One of the biggest challenges in weighting for causal effect estimation of functional treatments 
    is that, unlike discrete or continuous variables, the density of functional treatment is not well established due to its intrinsically infinite dimension \citep{delaigle2010defining}. 
    This phenomenon also prevents a direct adaptation of existing estimators for continuous treatment effects which often requires estimating density functions.  
    To overcome this issue, we define a weight function through reverse conditioning that is well-defined for functional treatments and properly adjusts for the selection bias.
    We also propose a novel estimation approach that directly computes weights via the idea of covariate balancing. Covariate balancing has recently become a popular approach in causal inference for observational studies due to its advantage of 
    providing a stable estimation of weights. 
    For example, covariate balancing methods have been developed for binary treatments \citep[e.g.,][]{hainmueller2012entropy,imai2014covariate, qin2007empirical,zubizarreta2015stable,wong2018kernel, wang2020minimal} and continuous treatments \citep[e.g.,][]{fong2018covariate,kallus2019kernel, tubbicke2022entropy}, and for estimating conditional treatment effects \citep[e.g.,][]{wang2022estimation}.

    \cite{zhang2021covariate} and \cite{tan2022causal} 
    also adopt the idea of covariate balancing to estimate the causal effect of functional treatments by weighting, but with several weaknesses.
    The approach by \cite{zhang2021covariate} 
    relies on a \emph{finite} approximation
    of the functional treatment by truncating the tail part of its functional principal components. The weights are estimated by balancing the correlation between the covariates and the top functional principal components, which is 
    directly generalized from \cite{fong2018covariate} to handle continuous treatments. 
    This approach has several drawbacks. First, there is likely information loss in selecting only several top functional principal components. Second, only balancing the \emph{correlation} may not be enough to guarantee the consistency of the final causal effect estimator unless the true outcome regression has a certain simple parametric form in the selected top functional principal components. Furthermore, the theoretical properties of their approach are not studied.
    Instead of using an approximation
    of the functional treatment, 
    \cite{tan2022causal} construct functional stabilized weights by balancing a set of growing number of basis functions. However, they focus on a functional linear marginal structural model, which imposes additional structure to the causal effect which may well be misspecified. %
    Compared with \cite{zhang2021covariate} and \cite{tan2022causal}, 
    our proposed covariate balancing method is distinct in the following aspects. (1) Our proposed method does not rely on any finite approximation of the functional treatment. (2) The balancing weights are constructed to directly balance the difference between the final causal effect estimator and the true target function. (3) We do not require a linear functional marginal structural model. (4) Our estimator attains the optimal rate of convergence under mild conditions.  We will further elaborate on these points as follows.
    
    Inspired by the development in nonparametric functional regression under the framework of reproducing kernel Hilbert space (RKHS) \citep[e.g.,][]{kadri2010nonlinear, zhang2012refinement, oliva2015fast, szabo2016learning, kadri2016operator}, we adopt the RKHS modeling for the functional treatment effects, which allows for a great flexibility (compared to a functional linear marginal structural model) in characterizing the effect of different functional treatments. In particular, we assume that the marginal structural model lies in an RKHS of functions with a functional input. 
    With the help of the closed-form solution of weight-modified kernel ridge regression (WMKRR),   
    we then propose balancing weights that are capable of controlling the balancing error between a smoothed weighted average and the population mean of functionals in a large hypothesis class 
    (see \eqref{eqn:bal1} for the explicit form).  We will then show that the solution to the optimization objective lies in a finite-dimensional space by a representer theorem and that 
    the resulting optimization is convex.  The theoretical analysis of the balancing error is not a trivial generalization from that for the binary treatment effects, since the balancing structure is significantly more complicated due to the interplay between weighting and smoothing in the formulation, and the balancing error is a function with a functional input instead of a scalar. Furthermore, while ignored by 
    \cite{zhang2021covariate} and \cite{tan2022causal}, a functional treatment is often not fully observable in practice and thus requires some pre-processing steps for recovery, which creates another layer of complication in the theoretical analysis. We provide a careful and detailed theoretical analysis to deal with all these complications. 
    Asymptotic properties of the proposed estimator are derived under the complex dependency
    structure of the weights and kernel ridge regression. Under appropriate technical conditions, we are able to  show that the proposed causal effect estimator can achieve the optimal nonparametric convergence rate, without additional modeling assumptions on the true weight function.

    The rest of the paper is organized as follows. Section \ref{sec:Setup} provides the basic setup of the weight-modified kernel ridge regression. Section \ref{sec:method} introduces the construction of covariate balancing weights for functional treatments. Computational details and an algorithm to construct the proposed balancing weights are presented in Section \ref{sec:computation}. Section \ref{sec:theory} develops the asymptotic properties of the proposed weighted estimator. The numerical performance of the proposed method is demonstrated by a simulation study in Section \ref{sec:simulation} and an application to a physical activity tracking data set in Section \ref{sec:real}.

    \section{Functional Treatment Effect Estimation}
    \subsection{Background and motivation}
    \label{sec:Setup}
    Let $A \in \mathcal{A}$ be a functional treatment defined on $\mathcal{T}$, and  $\bm X \in \mathcal{X} \subseteq \mathbb{R}^p$ be a $p$-dimensional confounder. 
    Denote by $Y(a)\in \mathbb{R}$ the potential outcome had treatment $a$ was given, for $a \in \mathcal{A}$.
    Suppose that \(\{(A_i, \bm X_i, Y_i(\cdot)): i = 1, \dots, n\}\) are independent and identically distributed copies of \((A, \bm X, Y(\cdot))\).  In practice, we do not observe all the potential outcomes per subject. In fact, only one particular case is observed.  The observed outcome for the $i$-th subject is $Y_i := Y_i(A_i)$.
    As such, the available data is
    \(\{(A_i, \bm X_i, Y_i): i = 1, \dots, n\}\). 
    The goal is to estimate the functional treatment effect (FTE): 
    \[\tau(a) : = \E \{Y(a)\}, \qquad a\in \mathcal{A}.\]
    Note that the domain of $\tau$ is a possibly infinite-dimensional function space $\mathcal{A}$, and therefore (non-parametric) estimations of $\tau$ are significantly harder than typical continuous treatment estimations. 
    Throughout the paper, we impose the following two assumptions. 
    \begin{assumption}[Weak unconfoundedness]
    Let $D(a)$ be the indicator of receiving treatment $a$: $D(a) = 1$ if $A = a$; $D(a) = 0$ otherwise. We have
        \label{assum:ignorability}
        $$Y(a) \ \ind\  D(a) \mid \bm X, \mathrm{\ for \  any\ } a \in \mathcal{A}.$$
    \end{assumption}
    This assumption is the weak unconfoundedness assumption introduced in \cite{imbens2000role}, which only requires the pairwise independence of the treatment with each of the potential outcomes. It is less restrictive than the strong ignorability assumption \citep{rosenbaum1983central}: $\{Y(a), a \in \mathcal{A}\}\  \ind\  A \mid \bm X$.
    
    Take $\rho_U$ as the marginal distribution of a random object $U$, and $\rho_{U\mid V}$ as the conditional distribution of $U$ given $V$. Define the weight function \(\truew\) as 
    \begin{align}
        \label{eqn:truew}
        \truew(a, \bm x) : = \frac{\rho_{\bm X}(\bm x)}{\rho_{\bm X\mid A}(\bm x\mid a)},  
    \end{align}
    which can be used to adjust the dependence between the treatment \(A\) and the confounder \(\bm X\). In \eqref{eqn:truew}, we use densities of the covariates $\bm X$ instead of the densities of the treatment $A$, as is commonly done in ATE methods \citep[e.g.,][]{wong2018kernel,kennedy2017non}. As such, we are able to circumvent the challenge of establishing functional densities \citep{delaigle2010defining}.
    Based on the definition of $\truew$, one can observe that
    \begin{align*}
        &\E \{\truew(A, \bm X) Y \mid A = a \}  = \E  \left[\truew(a, \bm X)\E( Y \mid A=a, \bm X ) 
     \mid A =a \right]  \\
      =  & \E  \left[\truew(a, \bm X)\E( Y(a) \mid A=a, \bm X ) 
     \mid A =a \right]
      =  \E  \left[\truew(a, \bm X)\E( Y(a) \mid D(a) = 1, \bm X ) 
     \mid A =a \right]\\
       =& \E  \left[\truew(a, \bm X)\E( Y(a) \mid  \bm X ) 
     \mid A =a \right]
       =   \E \left\{ \frac{\rho_{\bm X}(\bm X)}{\rho_{\bm X\mid A}(\bm X \mid a)} \E( Y(a) \mid  \bm X )  \mid A= a \right\} \\
        =  &\int_{\bm x}  \frac{\rho_{\bm X}(\bm x)}{\rho_{\bm X\mid A}(\bm x\mid a)} \E( Y(a) \mid  \bm X = \bm x )  \rho_{\bm X\mid A}(\bm x\mid a) \mathrm{d}{\bm x} 
        = \int_{\bm x}  \rho_{\bm X} (\bm x) \E( Y(a) \mid  \bm X = \bm x )  \mathrm{d}{\bm x} = \E \{ Y(a) \}. 
    \end{align*}
    Intuitively, this means that 
    the weight function $\truew$ helps to adjust the conditional expectation of the observed outcome $Y$ so that the weight-modified outcome $Z:=w^*(A,\bm X)Y$ is an \emph{unbiased} observation of the treatment effect. 
    This indicates that 
    one can regress $Z$ on $A$ to recover $\tau$.
    In addition to Assumption \ref{assum:ignorability}, we also require the following overlap condition:
    \begin{assumption}[Overlap]
        \label{assum:overlap}
        There exists a  positive constant \(\newu\ltxlabel{cnt:overlapupper}\) such that $\truew(a,\bm x) \leq \oldu{cnt:overlapupper}$ for all  \(a \in \mathcal{A}, \bm x \in \mathcal{X}\).
    \end{assumption}
    This assumption plays a similar role as the standard positivity assumption of the conditional treatment density under the settings of continuous treatments \citep[e.g., Assumption 2 in][]{kennedy2017non}. 
    If the true weights \(\truew_i := \truew(A_i, \bm X_i)\), \(i = 1, \dots, n\), are known and the functional treatments \(A_i\), $i=1,\dots,n$, are fully observed, we can construct the adjusted outcomes \(Z_i = \truew_i Y_i\) such that \(\E (Z_i\mid A_i) = \tau(A_i)\) for every \(i\),  and then perform a regression over the data \(\{(A_i, Z_i), i = 1,\dots, n\}\) to estimate $\tau$. Here we consider a nonparametric regression model that allows for a flexible modeling for \(\tau\). In particular, we assume that \(\tau\) lies in an RKHS \(\Hscr_A\) with a reproducing kernel \(\kernel_{A}(\cdot,\cdot)\), with the corresponding inner product 
    \(\langle \cdot, \cdot\rangle_{\Hscr_A}\) and norm \(\|\cdot\|_{\Hscr_A}\) respectively. 
    \begin{remark}
    \label{rmk:kernel_example}
        Note that  the reproducing kernel $\kernel_A$ is a bivariate function with function inputs.
        Based on an additional assumption that $\mathcal{A}$ is a Hilbert space, we provide some concrete examples of $\kernel_A$  that are easy to implement in practice. 
        Recall that its
        inner product and norm are denoted 
        by $\langle \cdot, \cdot\rangle_{\mathcal{A}}$ and $\|\cdot\|_{\mathcal{A}}$ respectively.
        The linear kernel $\kernel_A(A_1, A_2) =  \langle A_1, A_2 \rangle_{\mathcal{A}} + c $ with a constant $c$ 
        is probably the simplest example.
        Its corresponding RKHS $\mathcal{H}_A$ contains linear functions of the form $\beta_0+ \langle \beta_1,\cdot\rangle_{\mathcal{A}}$, where $\beta_0\in\mathbb{R}$ and $\beta_1\in\mathcal{A}$.
        As for nonlinear kernels, examples include the Gaussian kernel with 
        $\kernel_A(A_1, A_2) = \exp\{-2\|A_1 - A_2\|_{\mathcal{A}}^2/\theta\}$ 
        and exponential kernel with 
        $\kernel_A(A_1, A_2) = \exp\{-\|A_1 - A_2\|_{\mathcal{A}}/\theta\}$ respectively 
        for some pre-specified $\theta>0$.
    
     \end{remark}
     
    We consider the weight-modified kernel ridge regression (WMKRR) estimator for $\tau$:
    \begin{align}
        \label{eqn:tautilde}
        \tilde{\tau} := \argmin_{\tau \in \Hscr_A} \frac{1}{n}\sum_{i=1}^n \left( \truew_iY_i - \tau(A_i) \right)^2  + \lambda\|\tau\|^2_{\Hscr_A},
    \end{align}
    where $\lambda>0$ is a tuning parameter of the regularization.
    In \eqref{eqn:tautilde}, the norm $\|\cdot\|_{\Hscr_A}$ in the penalty term measures the ``roughness'' of the underlying mapping, and therefore encourages a ``smoother'' solution to \eqref{eqn:tautilde} as $\lambda$ increases.

    Unfortunately, the true weights \(\truew_i\), \(i = 1,\dots,n\), are typically unknown in observational studies. A natural solution is to first directly estimate \(\truew_i\) based on its definition, and then construct the adjusted outcomes based on these estimates. 
    This has been extensively studied in ATE estimation when the treatment \(A\) is a binary random variable \citep[e.g.,][]{feng2012generalized, hirano2003efficient}.
    However, this approach has several drawbacks.  First, from the definition of \(\truew\) in \eqref{eqn:truew}, the estimation of \(\truew\) involves estimating the densities of \(\bm X\) and \(\bm X\mid A\).
    Even if one uses a \emph{finite} approximation of \(A\) (see Remark \ref{rmk:fcbps} below), their estimations are still challenging. This is because to make Assumption \ref{assum:ignorability} plausible, 
    \(\bm X\) should include all the confounders that affect both the treatment and outcome, so it is usually multivariate. This indicates that  estimating multivariate density functions is required, which is known to be difficult:  
    Parametric estimations of multivariate density functions have a risk of possible model misspecifications, while nonparametric estimations such as the kernel density estimation suffer from slow rates of convergence in multivariate settings. 
     Second, the true weights are expected to achieve the balance for 
     covariates in expectation. However, it is unclear if such balance is enough for finite samples, especially when the sample size is small and the covariates are sparse \citep{zubizarreta2011matching}. 
    Third, the inverse of the densities can result in instability, especially when the estimated densities are close to zero.

    To overcome the aforementioned problems, we consider finding a stable set of weights that mimic the role of \(\truew_i\), \(i = 1,\dots, n\), through the idea of covariate balancing. 
    There exists extensive literature on covariate balancing techniques for ATE estimation when the treatment is binary \citep[e.g.][]{hainmueller2012entropy,imai2014covariate, qin2007empirical,zubizarreta2015stable,wong2018kernel, wang2020minimal} or continuous \citep[e.g.][]{fong2018covariate, kallus2019kernel,tubbicke2022entropy}, while we consider covariate balancing for the challenging setup where the treatment is functional.

    
    \begin{remark}
        \label{rmk:fcbps}
        \cite{zhang2021covariate} introduce functional propensity scores that are based on the functional principal components (FPCs) of \(A\) and define balancing weights based on the lower-order FPCs.  Then the input of the weight function becomes finite-dimensional, which resembles the setting of multivariate continuous treatments. 
        However, this definition relies on the unsupervised dimension reduction of the process $A$ and has the risk of missing important information. For example, if the higher-order FPC scores are more correlated with the potential outcome than the lower-order ones, 
        their proposed weight that only depends on the latter 
        may not properly account for all confounding.  In contrast, our method to be shown below 
        does not rely on such unsupervised truncation of $A$ so it can avoid the information loss mentioned above.    
    \end{remark}
    
    \begin{remark}
        \cite{tan2022causal} introduce a functional stabilized weight (FSW) estimator. They consider a functional linear marginal structural model $\tau(a) = \alpha  + \int_{\mathcal{T}} \beta(t) a(t) \mathrm{d}t$ for some scalar $\alpha$ and function $\beta$, which is restrictive and subject to the risk of model misspecifications.
        Instead of directly estimating the weight function $\truew$, they estimate its projection $\truew(a, \cdot): \mathcal{X} \rightarrow \mathbb{R}$ for every fixed $a \in \mathcal{A}$ by attempting to maintain the covariate balance $\E\{ \truew(a, \bm X) b(\bm X) \} = \E \{ b(\bm X)\}$ for any integrable function $b$.  A Nadaraya-Watson estimator is proposed to approximate the left-hand side of the equation. 
        To obtain the sequence of estimated weights, they have to perform the optimization $n$ times, separately for each observation.
        The convergence of their estimated weights depends on the smoothness of the projection $\truew(a, \cdot)$. 
        In contrast, as shown below, our method does not require the restrictive linearity assumption above for the marginal structural model. Moreover, our proposed weights are calculated jointly via a single optimization, and we do not require any smoothness assumption for the function $\truew$. Furthermore, the final weighted causal effect estimator can achieve the optimal rate of convergence with a nonparametric modeling of $\tau$, \emph{i.e.,} without assuming a linear functional marginal structural model on $\tau$.
    \end{remark}

    \subsection{Construction of weights}
    \label{sec:method}
    
    
    To motivate our construction, first suppose we have obtained a set of adjusted weights \( \bm w = [w_1,\dots, w_n]^\tp\). 
    From \eqref{eqn:tautilde}, we form an estimator of the treatment effect:
    \begin{align}
        \label{eqn:tauhat}
        \hat{\tau}_{\bm w} := \argmin_{\tau \in \Hscr_A} \frac{1}{n}\sum_{i=1}^n \left( w_i Y_i - \tau(A_i) \right)^2  + \lambda\|\tau\|^2_{\Hscr_A}.
    \end{align}
    Recall that \(\Hscr_A\) is the RKHS with the reproducing kernel \(\kernel_A\). We give the following definition which will be useful in expressing and analyzing the solution of \eqref{eqn:tauhat}.
    
    \begin{definition}
        \label{def:Kstar}
        For \(a \in \mathcal{A}\), \(\mathcal{K}_a: \mathbb{R} \rightarrow \Hscr_A\) is a Hilbert-Schmidt  operator such that 
        \[f(a) = \mathcal{K}_a^* f = \langle \kernel_A(a, \cdot), f \rangle_{\Hscr_A},\]
        where \(\mathcal{K}_a^*\) is the adjoint of \(\mathcal{K}_a\).
        Define the operator $\mathcal{S}_a :=\mathcal{K}_a  \mathcal{K}_a^*$. Note that we have
        \[\mathcal{S}_a: \Hscr_A \rightarrow \Hscr_A,  \qquad (\mathcal{S}_a f)(\cdot)= f(a) K_A(a,\cdot) \mathrm{\  for\  any \ } f \in \Hscr_A.\]
    \end{definition}

    With Definition \ref{def:Kstar}, 
    the estimator \eqref{eqn:tauhat} can be rewritten as:
    \begin{align}
        \label{eqn:tauw}
        \hat \tau_{\bm w}=  \left( \frac{1}{n}\sum_{i=1}^n \mathcal{S}_{A_i} + \lambda \mathcal{I} \right)^{-1}\left( \frac{1}{n}\sum_{i=1}^n \mathcal{K}_{A_i} w_i Y_i \right),
    \end{align}
    where $\mathcal{I}: \Hscr_A \rightarrow \Hscr_A$ is an identity operator such that $\mathcal{I}f = f$ for any $f \in \Hscr_A$.
    See, e.g., \cite{caponnetto2007optimal} and \cite{smale2007learning} for  more details.
    
    {Define $m(a, \bm x) = \E\{Y(a) \mid A=a, \bm X = \bm x\}=\E\{Y(a) \mid \bm X = \bm x\}$. We can then express
    \begin{align}
    \label{eqn:Y_i(a)}
        Y_i(a) = m(a, \bm X_i) + \epsilon_i(a), \qquad i = 1,\dots, n,
    \end{align}
    where $\epsilon_i(a) = Y_i(a) - m(a,\bm X_i)$ satisfies
    $\E[\epsilon_i(a) \mid A_i=a, \bm X_i] = \E[\epsilon_i(a) \mid \bm X_i] = 0$. This allows the error to be heteroskedastic with respect to the functional treatment and other covariates, and leads to $\tau(a)= \E_{X \sim \rho_X}\{m(a, X)\}$. 
    We assume that $\E [\epsilon^2_i(a) \mid \bm X_i] \leq \sigma_0^2 <\infty$ for some constant $\sigma_0>0$ (not depending on $\bm X_i$, $a$ and $i$).
    As $(Y_i(a), \bm X_i)$, $i=1,\dots,n$, are i.i.d.,
    so are $\epsilon_i(a)$, $i=1,\dots,n$.
    According to \eqref{eqn:Y_i(a)},  the observed data follow
    \begin{align}
    \label{eqn:Y_i}
        Y_i = Y_i(A_i) =  m(A_i, \bm X_i) + \epsilon_i(A_i)=
        m(A_i, \bm X_i) + \epsilon_i, \qquad i = 1,\dots, n,
    \end{align}
    where we write $\epsilon_i = \epsilon(A_i)$ for short.
    Clearly, $\E(\epsilon_i \mid A_i=a, \bm X_i) = \E(\epsilon_i(a) \mid A_i=a, \bm X_i)= \E(\epsilon_i(a) \mid \bm X_i) = 0$
    and, similarly, $\E(\epsilon_i^2 \mid A_i=a, \bm X_i) \le \sigma_0^2$. As such, $\E(\epsilon_i \mid A_i, \bm X_i) = 0$, $\E (\epsilon_i^2 \mid A_i, \bm X_i) \leq \sigma_0^2$.
    }
    Following \eqref{eqn:Y_i}, we can decompose the difference between \(\hat \tau_{\bm w}\) and \(\tau\)
    as:
    \begin{align}
      \hat \tau_{\bm w} - \tau =& \left( \frac{1}{n}\sum_{i=1}^n \mathcal{S}_{A_i} + \lambda \mathcal{I}\right)^{-1}\left( \frac{1}{n}\sum_{i=1}^n \mathcal{K}_{A_i} w_i Y_i\right) - \tau = I_1 + I_2, \nonumber\\
      \text{where} \quad I_1 &=\left( \frac{1}{n}\sum_{i=1}^n \mathcal{S}_{A_i} + \lambda \mathcal{I}\right)^{-1}\left( \frac{1}{n}\sum_{i=1}^n \mathcal{K}_{A_i} w_i m(A_i, \bm X_i)\right) - \tau, \label{eqn:decomp1}\\
      \text{and}\quad  I_2 & = \left( \frac{1}{n}\sum_{i=1}^n \mathcal{S}_{A_i} + \lambda \mathcal{I}\right)^{-1}\left( \frac{1}{n}\sum_{i=1}^n \mathcal{K}_{A_i} w_i \epsilon_i\right).  \label{eqn:decomp2}
    \end{align}
    Apparently the estimation error of $\hat\tau_{\bm w}$
    can be bounded by properly controlling the magnitudes of $I_1$ in \eqref{eqn:decomp1} and $I_2$ in \eqref{eqn:decomp2}.
    Roughly speaking, term \eqref{eqn:decomp2} exhibits concentration (at zero)
    due to the independence among $\epsilon_i$'s conditional on the treatments and covariates. This will be rigorously shown in our theoretical analysis.
    The primary challenge lies in controlling \eqref{eqn:decomp1} since $m$ is unknown in practice.
    To address this, motivated by \cite{wong2018kernel}, \cite{kallus2019kernel} and \cite{wang2022estimation}, we assume that \(m\) belongs to a certain class of functions and control \eqref{eqn:decomp1} for every element in this class. 
    
    Explicitly, we assume that \(m\) lies in a tensor-product RKHS \(\Hscr:=\Hscr_A \otimes \Hscr_X\) of functions defined on $\mathcal{A} \times \mathcal{X}$. 
    Here \(\Hscr_X\) is an RKHS of functions defined on $\mathcal{X}$, with 
    a reproducing kernel \(\kernel_X\) and its corresponding inner product
    and norm of $\Hscr_X$ are denoted by \(\langle \cdot, \cdot \rangle_{\Hscr_X}\) and \(\|\cdot\|_{\Hscr_X}\) respectively. Note that the assumption \(m \in \Hscr=\Hscr_A \otimes \Hscr_X\)
    is compatible with the aforementioned 
    assumption 
    $\tau\in\Hscr_A$
    due to the following proposition.
    \begin{proposition}
    \label{prop:tau}
    Under Assumption \ref{assum:kernelbound} in Section \ref{sec:theory}, if $\sup_{u \in \Hscr_X}|\E u(\bm X)|\neq 0$, we have $\Hscr_A=\{\E_{\bm X \sim \rho_X} g(\cdot, \bm X): g \in \Hscr\}$.
    \end{proposition}

    To bound the magnitude of \eqref{eqn:decomp1},  
    we aim to find 
    weights \(\tilde{\bm{w}} = [\testw_1, \dots, \testw_n]^{\tp}\) 
    such that 
    \begin{align} 
        \label{eqn:bal1}
    \Upsilon: = \sup_{u \in \Hscr: \|u\|_\Hscr \leq 1}\left\|\left( \frac{1}{n}\sum_{i=1}^n \mathcal{S}_{A_i} + \lambda \mathcal{I}\right)^{-1}\left( \frac{1}{n}\sum_{i=1}^n \mathcal{K}_{A_i} \testw_i u(A_i,  \bm  X_i)\right) -
    \E_{
    \bm X\sim \rho_{\bm X}}u(\cdot, \bm X)
    \right\|
    \end{align}
    is minimized with respect to some norm \(\|\cdot\|\). 
    Note that the objective (i.e., the norm) of the supremum \eqref{eqn:bal1} is proportional to $\|u\|_\mathcal{H}$.
    Therefore, we limit the space to $\Hscr(1)=\{u \in \Hscr: \|u\|_{\Hscr} \le 1\}$.
    Since \(\tau(a) = \E_{\bm X\sim \rho_{\bm X}} m(a, \bm X)\) and \(m \in \mathcal{H}\), we have 
    \begin{align*}
        \left\| \left( \frac{1}{n}\sum_{i=1}^n \mathcal{S}_{A_i} + \lambda \mathcal{I}\right)^{-1}\left( \frac{1}{n}\sum_{i=1}^n \mathcal{K}_{A_i} \tilde w_i m(A_i,\bm  X_i)\right) - \tau  \right\| \leq  \Upsilon \|m \|_{\Hscr}. 
    \end{align*}
    Since $\|m\|_\Hscr<\infty$, bounding \eqref{eqn:bal1} provides a good control over the discrepancy \eqref{eqn:decomp1} even if we do not know the true outcome function $m$.

    While $\bm\testw$ is well motivated, the criterion in \eqref{eqn:bal1} is not directly applicable for the following reasons.
    First, \(\E_{ \bm  X\sim \rho_{\bm X}}u(\cdot,  \bm  X)\) is usually unavailable since the distribution of \( \bm  X\) is unknown. Thus we propose to replace it with its empirical counterpart \(\sum_{i=1}^n u(\cdot,  \bm  X_i)/n\).  Second, the norm $\|\cdot\|$ in \eqref{eqn:bal1} needs to be chosen.
    A natural choice is $\mathcal{L}_2(A)$-norm $\|\cdot\|_{\mathcal{L}_2}$ defined by
    $\|f\|_{\mathcal{L}_2}=\sqrt{\E\{f^2(A)\}}$ for a function $f:\mathcal{A}\to\mathbb{R}$.
    In practice, we will use the empirical norm $\|\cdot\|_n$, which is defined by \(\|f\|_n : =\sqrt{ \sum_{k=1}^n f^2(A_k)/n}\).
    Finally, the functional treatments are often not fully observed in practice so we will need to recover $A_i$ by $\hat A_i$. 
    Two examples of $\hat A_i$ will be given in Examples \ref{exp:dense} and \ref{exp:kmm} below.

    By the above discussion, we 
    use the following criterion for controlling (\ref{eqn:decomp1}):
    \begin{align}
        \label{eqn:balerror}
        Q(\bm w,\lambda, u):=  \left\| \left( \frac{1}{n}\sum_{i=1}^n \mathcal{S}_{\hat A_i} + \lambda \mathcal{I}\right)^{-1}\left( \frac{1}{n}\sum_{i=1}^n \mathcal{K}_{\hat A_i} w_i u(\hat A_i,  \bm 
     X_i)\right)  - \frac{1}{n}\sum_{j=1}^n u(\cdot,  \bm  X_j) \right\|_n^2. 
    \end{align}
    In addition, to simultaneously control the second moment of \eqref{eqn:decomp2},
    we introduce the following regularization term of the weights:
    \begin{align}
        \label{eqn:penalty}
        R(\bm w,\lambda) =  \frac{1}{n^2} \sum_{i=1}^n w_i^2 \left\| \left( \frac{1}{n}\sum_{j=1}^n \mathcal{S}_{\hat A_j} + \lambda \mathcal{I} \right)^{-1}\mathcal{K}_{\hat A_i} \right\|_n^2.
    \end{align}
    Combining (\ref{eqn:balerror}) and (\ref{eqn:penalty}), we define the proposed balancing weights
    as 
    \begin{align}
        \bm \estw = [\estw_1,\dots, \estw_n]^\tp := \argmin_{0 \le w_i \le L, i = 1,\dots,n} \left[\sup_{u \in \Hscr(1)} \{Q(\bm w, \lambda, u)\}
         +  \eta R(\bm w, \lambda)\right], \label{eqn:estw_def}
    \end{align} 
    where $\Hscr(1)=\{u \in \Hscr: \|u\|_{\Hscr} \le 1\}$,  \(\eta\ge 0\)
    is a tuning parameter and \(L\) is an upper bound for the estimated weights and is allowed to be infinity. 
    In sequel, we write \(\hat{\tau}\) in short for \(\hat{\tau}_{\hat{\bm w}}\) when \(\hat{\bm w}\) is computed from \eqref{eqn:estw_def}.

    Before we conclude this section, we provide some examples of $\hat A_i$ to recover $A_i$ in practice, $i = 1,\dots, n$.  
    
    \begin{example}
    \label{exp:dense}
    \textbf{Densely observed trajectories}. For every $A_i \in \mathcal{A}$, $i=1,\ldots, n$, its noisy observations 
    $\{\gamma_{i,j}: j =1,\dots, N\}$ are measured at $\{t_{i,j}: j = 1,\dots,N\}$, a dense grid of $\mathcal{T}$. More specifically,  $\gamma_{i,j} = A_i(t_{i,j}) + \varepsilon_{i,j}$, where $\E (\varepsilon_{i,j}) =0$. %
    Under certain assumptions on the grid points, e.g., $t_{i,j}$, $j=1,\dots, N$, are i.i.d. copies of a random variable $T$ with density $\rho_T$, 
    common nonparametric regression procedures can be applied to each individual $i$ to obtain $\hat A_i$, such as 
    penalized spline regression \citep[e.g.][]{griggs2013penalized} and  smoothing spline regression \citep[e.g.][]{rice1983smoothing}.
    \end{example}

    \begin{example}
        \label{exp:kmm}
        \textbf{Kernel mean embedding of distributions}.
        If we want to use a distribution as a treatment (e.g., the activity profile as mentioned in Section \ref{sec:intro}), we can let $A$ be the kernel mean embedding of the distribution \citep[e.g.,][]{Muandet-Fukumizu-Sriperumbudur17}.
       More specifically, for the $i$-th individual,
       the kernel mean embedding $A_i$, $i=1,\ldots, n$, is defined by
    \begin{align*}
         A_i = \int_{\mathcal{T}} K_e(\cdot, t) \mathrm{d} P_i(t),
     \end{align*}
     where $P_i$ is the corresponding distribution function, 
    and $\kernel_e: \mathcal{T} \times \mathcal{T} \rightarrow \mathbb{R}$ is a reproducing kernel.
    Suppose that we have i.i.d.~samples $t_{i,l}$, $l=,\dots,N$, drawn from $P_i$. Then
    we can take $\hat A_i$ as the empirical embedding
     \begin{align*}
        \hat A_i =\frac{1}{N} \sum_{l=1}^N K_e(\cdot, t_{i,l}).
     \end{align*}
    \end{example}

    \section{Computational Details}
    \label{sec:computation}
    In this section, we discuss the computation for \eqref{eqn:estw_def}.
    
    \subsection{Representer theorem, closed-form expression and convexity}
    
    \subsubsection{Inner optimization}
    First, let us focus on the inner optimization of \eqref{eqn:estw_def},
    i.e., \(\sup_{u \in \Hscr(1)} Q(\bm w, \lambda, u)\).
    Note that it is an infinite-dimensional optimization problem
    when $\Hscr$ is infinite dimensional.
    The practical optimization relies on the following representer theorem,
    which shows that
    the solution indeed lies in a finite-dimensional space given data.
    
    \begin{thm}[Representer theorem]
    \label{thm:representer}
        The solution to $\sup_{u \in \Hscr(1)} Q(\bm w, \lambda, u)$
        lies in  the finite-dimensional space
    \begin{align}
        \Hscr_n := \left\{\sum_{i=1}^n \alpha_i \kernel_A(\cdot, \hat A_i) \kernel_X(\cdot, \bm X_i) + \sum_{i=1}^n \beta_i \kernel_A(\cdot, \hat A_i) \left(\frac{1}{n}\sum_{j=1}^n \kernel_X(\cdot, \bm  X_j)  \right) : \alpha_i, \beta_i \in \mathbb{R}\right\}. \nonumber
    \end{align}
    \end{thm}
    With Theorem \ref{thm:representer},
    we are able to take a further step and obtain a closed-form representation of  $\sup_{u \in \Hscr(1)} Q(\bm w, \lambda, u)$. Define
        \begin{align}
            \begin{split}
                \bm G_A := [\kernel_A(\hat  A_i, \hat  A_j)]_{i,j = 1}^n \in \mathbb{R}^{n\times n}; \qquad \bm G_X := [\kernel_X(\bm X_i, \bm X_j)]_{i,j = 1}^n \in \mathbb{R}^{n\times n};\\
             \bar{\bm G}_X := \left[ \frac{1}{n}\sum_{j=1}^n\kernel_X(\bm X_i, \bm X_j) \right]_{i=1}^n \in \mathbb{R}^n; \qquad  \bar{g}_X := \frac{1}{n^2} \sum_{i=1}^n \sum_{j=1}^n \kernel_X(\bm X_i, \bm X_j). 
            \end{split}
             \label{eqn:gram_matrices}
        \end{align}
    Then, for any $u \in \Hscr_n$, $Q(\bm w, \lambda, u)$ can be written as
    \begin{align*}
       Q(\bm w, \lambda,u)  
       =& \frac{1}{n} \big\|  \bm G_A \left( \bm G_A + n \lambda \bm I\right)^{-1} \left[ \bm w \circ \left\{ \left( \bm G_A \circ \bm G_X \right)\bm \alpha + \left(\bm G_A \odot \bar{\bm G}_X^\tp   \right)^\tp \bm \beta \right\} \right]  \\
       & -  \left[ \left(\bm G_A \odot \bar{\bm G}_X^\tp   \right) \bm \alpha +  \bar{g}_X \bm G_A \bm \beta\right] \big\|_2^2, \quad \text{for some  $\bm \alpha \in \mathbb{R}^n $ and $\bm \beta \in \mathbb{R}^n$,} 
    \end{align*}
    where \(\circ\) is the element-wise product between matrices (vectors),  \(\odot\) is the column-wise Khatri-Rao product, and $\|\cdot\|_2$ is the Euclidean norm of a vector. 
    
    We next simplify the constraint $u \in \Hscr(1)$ in maximizing $Q(\bm w, \lambda,u)$. By Theorem \ref{thm:representer}, it suffices to only focus on  
    the squared RKHS norm of a function \(u \in \Hscr_n\), which can be expressed as
    \begin{align}
        &\left\| \sum_{i=1}^n \alpha_i \kernel_A(\cdot, A_i) \kernel_X(\cdot, \bm  X_i) + \sum_{i=1}^n \beta_i \kernel_A(\cdot, A_i) \left(\frac{1}{n}\sum_{j=1}^n \kernel_X(\cdot, \bm  X_j)  \right) \right\|^2_\Hscr \nonumber 
        =\bm\gamma^\tp \bm G_F  \bm \gamma, \nonumber
    \end{align}
    where \(\bm \gamma = [\bm \alpha^\tp, \bm \beta^\tp]^\tp \in \mathbb{R}^{2n}\) and 
    \[\bm G_F = 
    \begin{bmatrix} 
     \bm G_A \circ \bm G_X &  \left(\bm G_A \odot \bar{\bm G}_X^\tp   \right)^\tp \\
     \bm G_A \odot \bar{\bm G}_X^\tp   & \bar{g}_X \bm G_A \\
      \end{bmatrix}
      \in\mathbb{R}^{2n\times 2n}.
    \]
    Note that we can decompose $\bm G_F$ as \[\bm G_F =\bm M \bm M^\tp= \begin{bmatrix} 
        \bm M_1 \\
        \bm M_2 \\
         \end{bmatrix}
         [\bm M_1^\tp, \bm M_2^\tp]
         ,\]
         where \(\bm M \in \mathbb{R}^{2n\times q}\), \(\bm M_1, \bm M_2 \in \mathbb{R}^{n \times q}\), \(q \leq 2n\).
         Thus finally, we have
    \begin{align}
        \sup_{u\in \Hscr(1)} Q(\bm w, \lambda, u)
         =& \frac{1}{n} \sup_{\bm \gamma^\tp \bm G_F \bm \gamma = 1} \left\| \bm G_A \left( \bm G_A + n \lambda \right)^{-1} \bm I \left\{ \mathrm{diag}(\bm w)\bm M_1 \bm M^\tp  \bm \gamma \right\} - \left( \bm M_2\bm M^\tp \right) \bm \gamma \right\|_2^2 \nonumber\\
         = & \frac{1}{n} \sup_{\|\bm M \bm \gamma\|_2 = 1}(\bm M^\tp \bm \gamma)^\tp \left\{ \bm G_A \left( \bm G_A + n \lambda \bm I\right)^{-1}   \mathrm{diag}(\bm w)\bm M_1 -  \bm M_2 \right\}^\tp  \nonumber \\
         & \qquad \left\{ \bm G_A \left( \bm G_A + n \lambda \right)^{-1}   \mathrm{diag}(\bm w)\bm M_1 -  \bm M_2 \right\} (\bm M^\tp \bm \gamma)\nonumber\\  
         = & \frac{1}{n} \left[ \sigma_{\max} \left\{ \bm G_A \left( \bm G_A + n \lambda \bm I\right)^{-1}   \mathrm{diag}(\bm w)\bm M_1 -  \bm M_2 \right\} \right]^2, \label{eqn:Q_expression}
    \end{align}
    where \(\sigma_{\max}(\cdot)\) returns the largest singular value of the input matrix. 
    As such, \eqref{eqn:Q_expression} is the closed-form representation of the objective function in the inner optimization of \eqref{eqn:estw_def}.
    
    \subsubsection{Convexity with respect to weights}

    We next show that the objective function in \eqref{eqn:estw_def} is convex with respect to $\bm w$. First, the regularization term $R(\bm w, \lambda)$ in the outer minimization is a quadratic function of $\bm w$ and hence convex in $\bm w$. Moreover, the inner maximization has been rewritten as in \eqref{eqn:Q_expression}, and its convexity in $\bm w$ is implied by the following lemma.

    \begin{lemma}
    \label{lem:convexity}
        For fixed $\bm A \in \mathbb{R}^{n\times n}$, $\bm B \in \mathbb{R}^{n\times q}$ and $\bm D \in \mathbb{R}^{n\times q}$, the function $ \varrho(\bm w) =  \left[ \sigma_{\max} \left\{ \bm A \mathrm{diag}(\bm w)\bm B -  \bm D \right\} \right]^2$ is a convex function.
    \end{lemma}

    Finally, we collect the previous results and express \eqref{eqn:estw_def} in a practical optimization form.
    Notice that
    \begin{align}
       R(\bm w, \lambda) = \frac{1}{n^2}\sum_{i=1}^n w_i^2\left\| \left( \frac{1}{n}\sum_{i=1}^n \mathcal{S}_{A_i} + \lambda \mathcal{I} \right)^{-1}\mathcal{K}_{A_i} \right\|_n^2 
         = \sum_{j=1}^n w_j^2 \left\{ \frac{1}{n}\sum_{i=1}^n \left\{ \left[ \bm G_A \left( \bm G_A + n \lambda \bm I \right)^{-1}  \right]_{i,j} \right\} ^2  \right\},
    \end{align}
    where $\bm A_{i,j}$ indicates the $(i,j)$-th element of matrix $\bm A$.
    Therefore we can show that
    \eqref{eqn:estw_def} is equivalent to 
    \begin{align}
        \begin{split}
            \hat {\bm w} =\argmin_{0 < w_i < L, i = 1,\dots,n}  \frac{1}{n}  \left[ \sigma_{\max} \left\{ \bm G_A \left( \bm G_A + n \lambda \bm I\right)^{-1}   \mathrm{diag}(\bm w)\bm M_1 -  \bm M_2 \right\} \right]^2\\
            + \eta \sum_{j=1}^n w_j^2 \left\{ \frac{1}{n}\sum_{i=1}^n \left\{ \left[ \bm G_A \left( \bm G_A + n \lambda \bm I \right)^{-1}  \right]_{i,j} \right\} ^2  \right\}.
        \end{split}
        \label{eqn:estw_opt} 
    \end{align}
    Due to the convexity of the objective function, common algorithms such L-BFGS-B can be applied to solve \eqref{eqn:estw_opt} given the smoothing parameter $\lambda$ and tuning parameter $\eta$.

    \subsection{Tuning parameter selection}
    \label{sec:tuning}
    
    Here we discuss how to select \(\lambda\) and \(\eta\). The smoothing parameter $\lambda$ needs to be provided in order to calculate the balancing error \eqref{eqn:balerror}, and hence the weights.
    Recall that the weights are used to form a modified outcome for the WMKRR. Naturally, one would tune $\lambda$ based on common methods for kernel ridge regression such as cross-validation, 
    but this becomes 
    very difficult in our case because of the complicated dependency between the weights and $\lambda$.
    To address this issue, we propose a simple solution which performs reasonably well in practice. The idea is to use a simple estimator of the adjusted response to guide the selection of $\lambda$. 
    More specifically, we first obtain the adjusted responses with the \fcbps{} weights described in \cite{zhang2021covariate}, as their weights do not depend on $\lambda$ and can be computed quickly. Then 
    we apply the leave-one-out cross-validation (LOOCV) to select $\lambda$ based on the mean square error computed in the validation set. 
    Finally we use the selected $\lambda$ to compute the proposed weights without updating $\lambda$ further.

    As for the hyper-parameter $\eta$, it is related to the magnitude of weights so as to achieve a balance between \eqref{eqn:decomp1} and \eqref{eqn:decomp2}. Here we propose to use a fitted outcome regression to help select the best $\eta$. To be specific, we fit a KRR to get an estimate for $m$, and denote it as $\hat m$. Then we take $\hat \tau_{\reg} = \frac{1}{n} \sum_{i=1}^n\hat m(\cdot, \bm  X_i)$ as the estimator for $\tau$ from the regression approach. Denote by $\hat w_i^{(\eta)}$, $i = 1,\dots, n$, the weights defined in \eqref{eqn:estw_def}
    for each given $\eta$. 
    We select the best $\eta$ such that 
    \begin{multline}
        \label{eqn:eta_tune}
         V(\eta): =  \left\| \left( \frac{1}{n}\sum_{i=1}^n \mathcal{S}_{\hat A_i} + \lambda \mathcal{I}\right)^{-1}\left( \frac{1}{n}\sum_{i=1}^n \mathcal{K}_{\hat A_i} \hat w^{(\eta)}_i \hat m(\hat A_i, \bm X_i)\right)  - \hat \tau_{\reg} \right\|_n^2  \\
       +   \left\| \left( \frac{1}{n}\sum_{i=1}^n \mathcal{S}_{\hat A_i} + \lambda \mathcal{I} \right)^{-1}\left[ \frac{1}{n}\sum_{i=1}^n \mathcal{K}_{\hat A_i} \hat w^{(\eta)}_i\left\{ Y_i - \hat m(\hat A_i, \bm X_i)\right\}\right] \right\|_n^2   
    \end{multline}
    is the smallest.  
    In Algorithm \ref{algo:tuning} in Section \ref{sec:algorithm_table} in the supplementary material, we summarize the computation steps to obtain the proposed $\hat \eta$ with tuning parameter selection.

    \section{Theory}
    \label{sec:theory}
    
    In this section we provide the rate of convergence for $\hat{\tau}$. We first introduce a few notations and assumptions.
    Given two Hilbert spaces $\mathcal{V}$ and $\mathcal{W}$, the operator norm  of an operator $B \in \mathcal{L}(\mathcal{V},\mathcal{W})$ is defined as
    $\| B \|_{\mathcal{L}(\mathcal{V},\mathcal{W})} : = \sup_{\|v\|_{\mathcal{V}} \leq 1} \|B v\|_{\mathcal{W}}$.

    \begin{assumption}
      \label{assum:errors}
      $\{(A_i, \bm X_i, \epsilon_i(a), a \in \mathcal{A}), i = 1,\dots, n \}$ are independent. For any $a \in \mathcal{A}$,
     $\E(\epsilon_i(a)|A_i, \bm X_i)=0$ for $i =1,\dots,n$, and $\sup_{1 \leq i \leq n}\E(\epsilon_i(a)^2|A_i, \bm X_i)\le \sigma^2_0$ for some constant $\sigma_0 >0$. 
    \end{assumption}

    \begin{assumption}
      \label{assum:functionclass}
      \(\tau\in \Hscr_A\) and \(m \in \Hscr\). 
    \end{assumption}

    \begin{assumption}
        \label{assum:kernelbound}
       $K_X$ and $K_A$ are positive definite kernels. $K_X$ is continuous and the real function 
       $(a_1, a_2) \mapsto \langle \mathcal{K}_{a_1} c_1, \mathcal{K}_{a_2} c_2\rangle_{\Hscr_A}$ is measurable for any $c_1, c_2 \in \mathbb{R}$.
       There exists a constant \(\newu\ltxlabel{cnt:kernelbound}>0\) such that $\sup_{a \in \mathcal{A}} |\kernel_A(a,a)|\le C_2$ and $\sup_{\bm x\in \mathcal{X}}|\kernel_X(\bm x, \bm x)| \}\leq \oldu{cnt:kernelbound}$. 
    \end{assumption}
    
    \begin{assumption}
    Either one of the following two conditions is satisfied. 
    \begin{enumerate}
      \label{assume:function_approximation}
        \item[(a)] Functional treatments $A_i$, $i = 1, \dots, n$, are fully observed without
        error. In this case, we take $\hat A_i = A_i$. This corresponds to $\kappa = 0$ in \eqref{eqn:kappa} below.
        \item [(b)]      
         There exists a pseudometric $d:\mathcal{A}\times\mathcal{A}\to [0,\infty)$ such that:
        \begin{enumerate}
            \item[(i)] The mapping $\mathcal{K}_{(\cdot)}:  \mathcal{A} \rightarrow \mathcal{L}(\mathbb{R}, \Hscr_A)$ is H\"{o}lder continuous, i.e.,
            there exist constants $H>0$ and $0<h\le 1$ such that
        \[\|\mathcal{K}_{a_1} - \mathcal{K}_{a_2}\|_{\mathcal{L}(\mathbb{R}, \Hscr_A)} \leq H [d(a_1, a_2)]^h, \quad a_1,a_2\in\mathcal{A}.
        \]  
        \item[(ii)] All $\hat{A}_1,\dots,\hat{A}_n$ are independent. Each $\hat{A}_i$ can estimate $A_i$ at a uniform rate $\kappa=\kappa(n)$ over $i=1,\dots,n$, i.e.,
        \begin{equation}
        \label{eqn:kappa}
            \max_{1\le i\le n} d(\hat A_i,A_i) = \bigOp \left( \kappa  \right).
        \end{equation}

        \end{enumerate}
    
    \end{enumerate}

    \end{assumption}

    Assumption \ref{assum:errors} is a standard assumption for the data, which allows for heteroscedasticity of the errors. Assumption \ref{assum:functionclass} states that the function classes for the target function \(\tau\) and the outcome model \(m\) are well-specified.
    Assumption \ref{assum:kernelbound} is a boundedness requirement for the reproducing kernels. It is satisfied for the majority of common kernels including Gaussian and exponential kernels discussed in Remark \ref{rmk:kernel_example}.
    As for Assumption \ref{assume:function_approximation},
    we only need one of the two specified conditions.
    When all functional treatments are fully observed without error (Assumption \ref{assume:function_approximation}(a)),
    there is no need to recover $A_i$. So we can simply take $\hat{A}_i=A_i$. 
    Otherwise, we need to recover $A_i$ by $\hat{A}_i$. 
    Assumption \ref{assume:function_approximation}(b) specifies the related conditions for 
    $\hat{A}_i$ in this case: 
    the H\"{o}lder continuity of the operator $\mathcal{K}_a$ and rate of convergence for $\hat A_i$.
    For example, 
    when we take $d$ in Assumption \ref{assume:function_approximation}(b) as the norm $\|\cdot\|$ 
     used in constructing the kernel in Remark \ref{rmk:kernel_example}, 
    the Gaussian kernel and exponential kernel mentioned in Remark \ref{rmk:kernel_example} satisfy the H\"{o}lder continuity condition with $h=1$ and $h = 1/2$ respectively. See Table 1 in \cite{szabo2016learning} for more examples. 
     As for the rate of convergence $\kappa$, it can be specified given different applications.
    As in Example \ref{exp:dense} in Section \ref{sec:method}, if every $A_i$, $i=1,\ldots, n$, satisfies 
     $\int_\mathcal{T} A_i^2(t) \rho_T(t) \mathrm{d}t < \infty$ ($\rho_T$ was defined as in Example \ref{exp:dense}), one can fix the norm
     $\|f\| = [\int_{\mathcal{T}} f^2(t) \rho_T(t) \mathrm{d}t]^{1/2}$ for each $f \in \mathcal{A}$ and obtain a nonparametric convergence rate for $\kappa$ with typical nonparametric regression approaches. For example, when $A_i$ is a twice-differentiable univariate function, under appropriate assumptions, the smoothing spline regression can lead to $\kappa = N^{-2/5} \log n$ \citep{raskutti2012minimax}, where $\log n$ is due to the union bound over $n$ functions. 
     In Example \ref{exp:kmm} in Section \ref{sec:method}, when $\hat A_i$, $i=1,\ldots, n$, are empirical kernel embeddings, one can take $\|\cdot\|= \|\cdot\|_{\Hscr_e}$, the RKHS norm of the kernel embeddings associated with the reproducing kernel $K_e$, 
     and let $\kappa = N^{-1/2} \log n $. 
    See Section A.1.10 in \cite{szabo2015two} for more detailed results.

    We introduce additional terms before presenting the last technical assumption. 
    Define $\mathcal{S}_a = \mathcal{K}_a \mathcal{K}_a^*$ and 
    let \(\mathcal{S} = \E_{A\sim \rho_A} \mathcal{S}_{A}\). 
    Define the trace norm $\mathrm{Tr}(\cdot)$ of a semi-positive-definite operator $B: \mathcal{B} \rightarrow \mathcal{B}$ as   $\mathrm{Tr}(B) = \sum_{l}\langle B e^\mathcal{B}_l, e^{\mathcal{B}}_l  \rangle_{\mathcal{B}}$ with $\{e^{\mathcal{B}}_l\}$ an  orthonormal basis of $\mathcal{B}$.
    Under Assumption \ref{assum:kernelbound}, \(\mathrm{Tr}(\mathcal{S} ) \leq \sup_{a \in \mathcal{A}} \mathrm{Tr}(\mathcal{S}_a) \leq \oldu{cnt:kernelbound}\).    Then the spectral theorem yields 
    \begin{align}
        \mathcal{S} = \sum_{l =1}^L t_l \langle \cdot, e_l \rangle_{\Hscr_A} e_l,
    \end{align}
    where $\{e_l\}_{l=1}^L \subset \mathcal{\Hscr_A}$ such that $\langle e_l, e_{l'}\rangle_{\Hscr_A} =1$ if $l=l'$ and 0 otherwise, and \(t_1 \ge t_2 \ge \dots \ge t_L >0\), with \(\sum_{l=1}^L t_l = \mathrm{Tr}(\mathcal{S}) \leq \oldu{cnt:kernelbound}\). Here \(L\) can be \(\infty\). 
    We also define 
      \[
      \mathcal{N}(\lambda) :=  \mathrm{Tr}\{(\mathcal{S} + \lambda\mathcal{I})^{-1}\mathcal{S}\} = \sum_{l=1}^\infty \frac{t_l}{t_l + \lambda}.
      \]

    As to be shown in the theorems below, 
    the rate of convergence for $\hat{\tau}$ depends on the decay of the eigenvalues of \(\mathcal{S}\). We also note that the theorems below hold for general choices of kernels.

    \begin{thm}
      \label{thm:emp_bal_error_truew}
      Under Assumptions \ref{assum:ignorability}--\ref{assume:function_approximation}, if \(\mathcal{N}(\lambda)(\lambda n)^{-1} = \bigO(1)\), \(\lambda \leq \|\mathcal{S}\|_{\mathcal{L}(\Hscr_A)}\), 
       $\kappa^{2h}  = \bigO(\lambda \mathcal{N}(\lambda) n^{-1})$, $\kappa^h = \bigO(\lambda)$ and \(\sqrt{ \sum_{l=1}^\infty \min \{t_l,\lambda\}}(\sqrt{n}\lambda)^{-1} = \bigO(1)\), we have 
      \begin{align}
        \label{eqn:emp_Qwstarl2bound}
          \sup_{u\in \Hscr(1)}Q(\hat{\bm{w}},\lambda,  u)  = \bigOp\left[ (1+ \eta)\frac{\mathcal{N}(\lambda)}{n} + \lambda    \right],
      \end{align}
      \begin{align}
        \label{eqn:emp_wstarpenalty}
       \text{and} \quad 
        R(\hat{\bm{w}},\lambda)= \bigOp\left[\frac{\mathcal{N}(\lambda)}{n} + \eta^{-1} \left(  \frac{\mathcal{N}(\lambda)}{n} + \lambda \right) \right],
      \end{align}
      where $\hat{\bm w}:=(\hat w_1,\dots,\hat w_n)^\tp$.
      \end{thm}
    
    Theorem \ref{thm:emp_bal_error_truew} provides the orders of the balancing error \eqref{eqn:balerror} and the regularization \eqref{eqn:penalty} with respect to
    $\hat{\bm w}$. 
    Based on Theorem \ref{thm:emp_bal_error_truew}, we can develop the rate of convergence for the proposed weighted estimator $\hat \tau$. \modify{For instance, if the decay of eigenvalues of $\mathcal{S}$ follows a polynomial rate as shown in Assumption \ref{assum:eigen_decay} below, we are able to achieve a nonparametric convergence rate for $\tau$.}

    \begin{assumption}
      \label{assum:eigen_decay}
      There exists a constant $b>1$ such that  $ t_l \asymp l^{-b}$ for any $l\ge 1$. 
    \end{assumption}

    This decay rate is also considered in \cite{caponnetto2007optimal} and \cite{szabo2016learning}.

      \begin{thm}
        \label{thm:emp_what_regress}
    \modify{Suppose that the conditions  
    stated in Theorem \ref{thm:emp_bal_error_truew}
    hold. Then
    \begin{align*}
        \|\hat \tau - \tau\|_n = \bigOp\left( \sigma_0 \eta^{-1/2} \left( \frac{\mathcal{N}(\lambda)}{n} + \lambda  \right) ^{1/2} + \left( \frac{\mathcal{N}(\lambda)}{n}  \right)  ^{1/2} \right) \|m\|_{\mathcal{H}}.
    \end{align*}
    If we further assume Assumption \ref{assum:eigen_decay} hold, 

        $\lambda \asymp n^{-b/(1+b)}$, $\| \mathcal{S}\|_{\mathcal{L}(\Hscr_A)} \ge \lambda$, $\kappa = \bigO(n^{-b/[h(1+b)]})$ and $\eta \asymp 1$, then 
        there exists some constant $C_b>0$ such that $$\mathcal{N}(\lambda) \leq C_b \lambda^{-1/b}.$$ 
        Also, we have
         \begin{align*}
          \left\| \hat \tau - \tau \right\|_n = \bigOp\left( n^{-\frac{b}{2(1+b)}} \right) \|m\|_{\Hscr}.
        \end{align*}
        }

        \end{thm}
    Theorem \ref{thm:emp_what_regress} provides the rate of convergence for $\hat \tau$ \modify{under Assumption \ref{assum:eigen_decay}}.
    According to \cite{caponnetto2007optimal}, this  rate is 
    minimax 
    in 
    estimating the target function $\tau \in \Hscr_A$ using the i.i.d data $\{A_i, \truew_i Y_i\}_{i=1}^n$. 
    
    \begin{remark}
          The proposed causal effect estimator, based on $\bm \estw$, enjoys the same minimax rate of convergence as the ordinary kernel ridge regression estimator using the modified outcome $\{\truew_i Y_i\}_{i=1}^n$ with the true but unknown weights.  However, 
          the theoretical analysis with weights obtained by \eqref{eqn:estw_def} is significantly more complicated than the typical analysis for kernel ridge regression. 
          One reason is that the responses in kernel ridge regression are typically assumed independent, while the adjusted responses $\estw_i Y_i$ are all dependent since $\bm \estw$ are obtained by \eqref{eqn:estw_def}. Moreover, as we do not impose any modeling assumption of $\bm \truew$, the convergence between $\bm \estw$ and $\bm \truew$ cannot be established or used to show the convergence of WMKRR. Instead we perform a careful analysis to control the uniform error $\sup_{u \in \Hscr(1)} Q(\bm \estw, \lambda, u) $, which leads to the convergence for $\hat \tau$.
    \end{remark}

    \section{Numerical Studies}

    \subsection{Simulation}
    \label{sec:simulation}
    We first compare the finite-sample performance of different estimators via a simulation study.
    We have $200$ simulated datasets where $n=200$ independent subjects are generated in each simulated data. The observations $(A_i, {\bm  X_i}, Y_i)$ for the $i$-th subject,  $i=1,\ldots, n$, in each simulated data are i.i.d.~copies of $(A, {\bm  X}, Y)$ as below. We assume the functional variables $A_i$'s are fully observed. 
    The confounders \(\bm  X = [X^{(1)}, X^{(2)}, X^{(3)}, X^{(4)}]^{\tp}\in \mathbb{R}^4\)
    follow the multivariate normal distribution with the zero mean vector and the identity covariance matrix. The functional treatment \(A\) is generated by \(A(t) = \sum_{k=1}^4 A^{(k)} \sqrt{2}\sin(2\pi k t)\), $t\in[0,1]$,
    where \(A^{(1)} \mid  \bm X \sim N(4X^{(1)}, 1)\), \(A^{(2)} \mid  \bm X \sim N(2\sqrt{3} X^{(2)},1)\), \(A^{(3)} \mid  \bm X \sim N(2\sqrt{2}X^{(3)}, 1)\) and \(A^{(4)} \mid \bm  X \sim N(2X^{(4)}, 1)\).

    The outcome \(Y\) is generated by \(Y \mid (A, \bm X) \sim N(m(A,\bm  X),1)\), 
    where we consider three choices for \(m\) as follows. 
    Let \(\Psi(\bm x) = x^{(2)}(x^{(1)})^2 + (x^{(4)})^2\sin(2x^{(3)})\) where $\bm x = [x^{(1)}, x^{(2)}, x^{(3)}, x^{(4)}]^{\tp}$ and \(\mu(t) = 2\sqrt{2}\sin(2\pi t) + \sqrt{2} \cos(2\pi t) + \sqrt{2}\sin(4\pi t)/2 + \sqrt{2}\cos(4\pi t) /2\).
    \begin{itemize}
        \item Setting 1: We let $m(a,\bm  x) = 15\Psi(\bm x) + \int_{t=0}^1 a(t) \mu(t) \mathrm{d}t.$
     In this case, the treatment effect $\tau(a)$ is linear in $a$ in the sense \[\tau(a) = \int_{t=0}^1 a(t) \mu(t) \mathrm{d}t, \] 
    and $m$ is additive in \(\Psi(\bm{x})\) and \(a\).
    \item Setting 2: We let $m(a,\bm x) = 10\Psi(\bm x) + 0.5(a^{(1)})^2 + 4\sin(a^{(1)}).$ Here $m$ is additive in \(\Psi(\bm{x})\) and \(a\).
    Then the treatment effect 
    is
    \[\tau(a)= 0.5(a^{(1)})^2 + 4\sin(a^{(1)}),\] 
    which is nonlinear in \(a\). 
    \item Setting 3: 
    We let $m(a,\bm x) = [1 + 2/3\Psi(\bm x)] [0.5(a^{(1)})^2 + 4\sin(a^{(1)})].$
    In this case, the treatment effect \(\tau(a)\) has the same form as in Setting 3, 
    but \(\Psi(\bm x)\) interacts with \(a\) in \(m\).
    \end{itemize}

    In this simulation study, we compare the following estimators for \(\tau\).  
    \begin{enumerate} 
       \item \proposed: our proposed (KRR) estimator where weights are obtained from \eqref{eqn:estw_def}.
        \item \fcbps: 
        \modify{the weighted least squares estimator proposed in \cite{zhang2021covariate} where the weights are based on the parametric SFPS estimation described in \citet{zhang2021covariate}. Here, the parametric modeling of the relevant densities is correctly specified. }
        \item \npfcbps: 
        \modify{the weighted least squares estimator proposed in \cite{zhang2021covariate} where the weights are obtained by using the non-parametric SFPS estimation described in \cite{zhang2021covariate}.}
     \item \reg: the regression estimator $\hat \tau_{\reg{}}$ discussed in Section 
     \ref{sec:tuning}. 
        \item \nw: the estimator without adjusting the response, i.e., \(w_i^* Y_i\) in \eqref{eqn:tautilde} is replaced by the original response \(Y_i\) for \(i = 1,\dots,n\). 
    \end{enumerate}
    When performing the KRR for  \modify{\proposed{}, \reg{} and \nw{}}, 
    we take \(\kernel_A\) and \(\kernel_X\)  both as  Gaussian kernels. More specifically,  \(\kernel_A(a_1, a_2) = (\sqrt{2\pi}\sigma_A)^{-1}\exp\{-\int_{t = 0}^1 (a_1(t) - a_2(t))^2 \mathrm{d}t/ \sigma_A^2\}\)
    and \(\kernel_X( \bm  x_1,  \bm  x_2) = (\sqrt{2\pi}\sigma_X)^{-1}\exp\{-(\bm x_1 - \bm x_2)^\tp (\bm x_1 - \bm x_2)/ \sigma_X^2\}\),
    where \(\sigma_A\) and \(\sigma_X\) are selected by the median heuristic \citep{fukumizu2009kernel, garreau2017large}.  For \fcbps{} and \npfcbps{},  the number of FPC \(L\) is chosen such that the top \(L\) FPC scores explain \(95\%\) percentage of the variance. \modify{Both \fcbps{} and \npfcbps{} assume a linear model for $\tau$. Therefore only Setting 1 is correctly specified for them. }
    For \proposed{}, we select its tuning parameter by the procedure described in Section \ref{sec:computation} (or Algorithm \ref{algo:tuning} in the supplementary material) 
    For \modify{\reg{} and \nw{}, } 
    we use LOOCV to select the smoothing parameter $\lambda$ in KRR.

    Two evaluation metrics are provided to assess the performance of these estimators. Take \(\tau'\) as a  generic estimator of \(\tau\).
    \begin{enumerate}
        \item Empirical MSE: the mean squared errors (MSE) on sample points: \(\sum_{i=1}^n [\tau(A_i) - {\tau'}(A_i)]^2/n\). 
        \item Out-of-Sample MSE: the mean squared errors (MSE) measured on a set of new evaluation points: \(\sum_{i=1}^{n'} [\tau(A'_i) - {\tau'}(A'_i)]^2/{n'}\), where \(n' =100\) and \(A_i'(t) = \sum_{k=1}^4 A_i'^{(k)} \sqrt{2}\sin(2\pi k t)\) with $A_i'$ sampled from the marginal distribution of $A_i$, i.e., \(A_i'^{(1)} \sim N(0, 17)\), \(A_i'^{(2)} \sim N(0, 13)\), \(A_i'^{(3)} \sim N(0, 9)\) and \(A_i'^{(4)} \sim N(0, 5)\).
    \end{enumerate}

    \modify{Tables \ref{tab:emp} and \ref{tab:eval} show the empirical MSEs and Out-of-Sample MSEs for the above estimators based on 200 simulated datasets respectively.}
    For both evaluation metrics, \nw{} has the worst performance among all settings, as it does not adjust for selection bias.
    \modify{\fcbps{} and \npfcbps{} perform worse than \proposed{} and \reg{} in Settings 2 and 3 as the assumption of linear model is violated in these two settings. Even though Setting 1 satisfies the linear assumption, the weights calculated from \fcbps{} and \npfcbps{} are only able to balance the case where the outcome model $m$ is linear in both $a$ and $x$. Therefore, they do not perform as well as \proposed{}.}
    Overall, \proposed{} achieves the smallest average of MSEs 
    among all five estimators 
    except for Setting 2 where it is outperformed only by \reg{}. 
    Moreover, the MSE associated with
    \proposed{} has the smallest standard errors in all three settings, which demonstrates its attractive stability.

    \begin{table}[h]
      \caption{Empirical MSEs for different estimators under three different simulation settings. Values in the parentheses are the standard errors of MSEs. }
      \label{tab:emp}
      \centering
      \begin{tabular}{r| lll}
        \hline
       & Setting 1 & Setting 2 & Setting 3 \\ 
        \hline
        \nw{} & 666.78 (21.43)& 306.06 (9.45) & 2687.51 (259.31) \\ 
        \fcbps{}  & 92.91 (2.83)  & 201 (3.48) & 194.57 (4.46) \\ 
        \npfcbps{} & 105.17 (2.44)  & 210.44 (3.44)& 228.26 (5.59) \\ 
        \proposed{}  & 34.82 (1.02) & 130.54 (2.77) & 138.15 (4.10) \\ 
        \reg{} & 94.22 (4.46) & 99.78 (4.11) & 182.23 (8.56) \\ 
         \hline
      \end{tabular}
      \end{table}

    \begin{table}[h]
      \centering
      \caption{Out-of-Sample MSEs for different estimators under three different simulation settings. Values in the parentheses are the standard errors of MSEs.}
      \label{tab:eval}
      \begin{tabular}{r| lll}
        \hline
       & Setting 1 & Setting 2 & Setting 3 \\ 
        \hline
      \nw{} & 516.04 (17.29) &  252.73 (9.17) & 1287.87 (92.64) \\ 
       \fcbps{} & 91.86 (2.82) & 205.00 (4.62) & 198.17 (5.11) \\ 
       \npfcbps{} & 105.39 (2.56) & 215.78 (4.62) & 231.27 (5.70) \\ 
      \proposed{} & 34.48 (1.03) &  133.41 (3.93) & 137.87 (4.31) \\ 
        \reg{} & 95.69 (4.80) & 105.25 (4.26) & 172.98 (7.60) \\ 
         \hline
      \end{tabular}
      \end{table}

    \subsection{Real Data Application}
    \label{sec:real}
\begin{figure}
      \centering
      \begin{subfigure}[t]{.5\linewidth}
        \centering\includegraphics[width=1\linewidth]{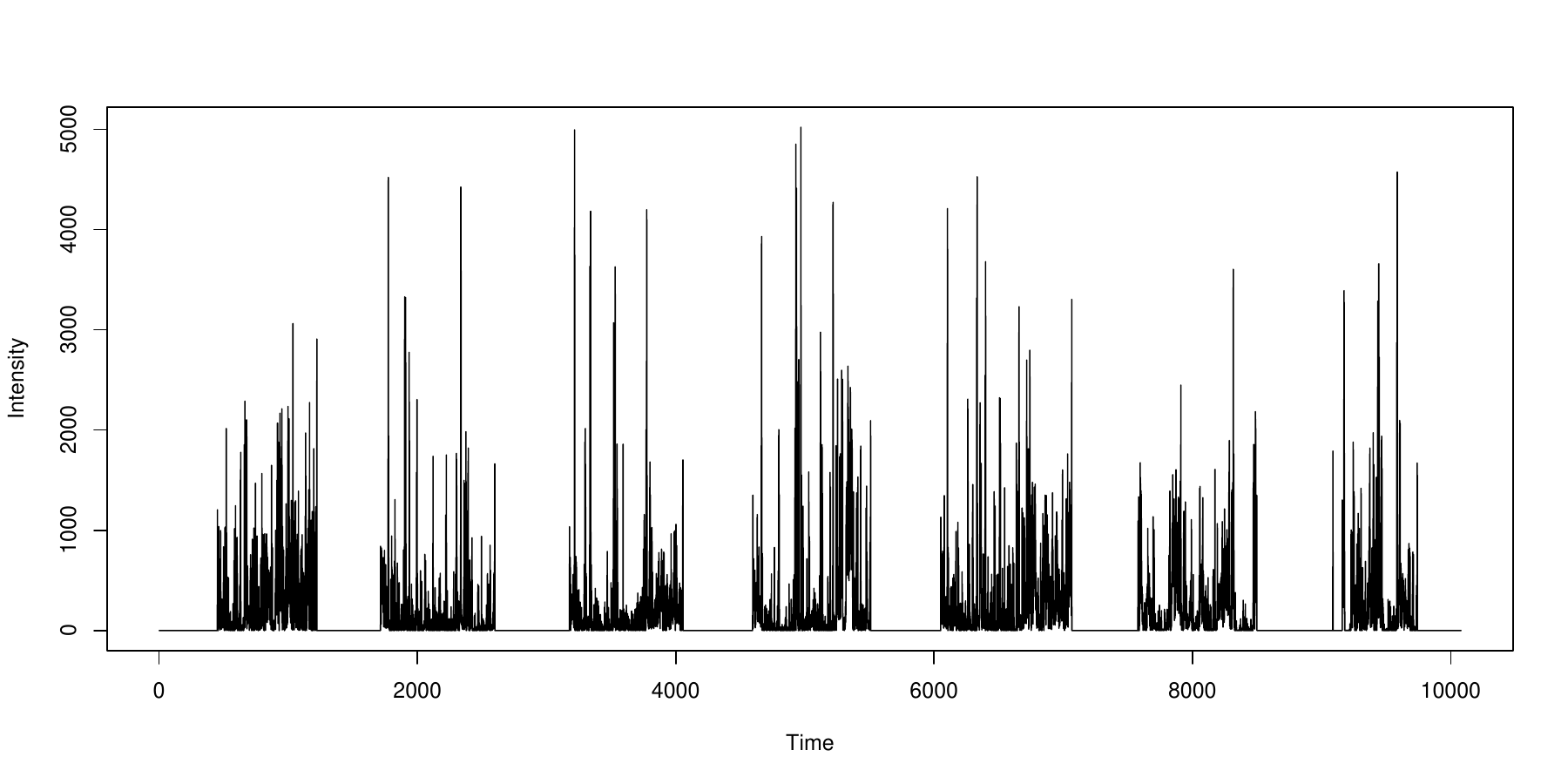}
        \caption{The trajectory of activity intensity values.}
          \label{fig:traj}
      \end{subfigure}
      \begin{subfigure}[t]{.4\linewidth}
        \centering\includegraphics[width=1\linewidth]{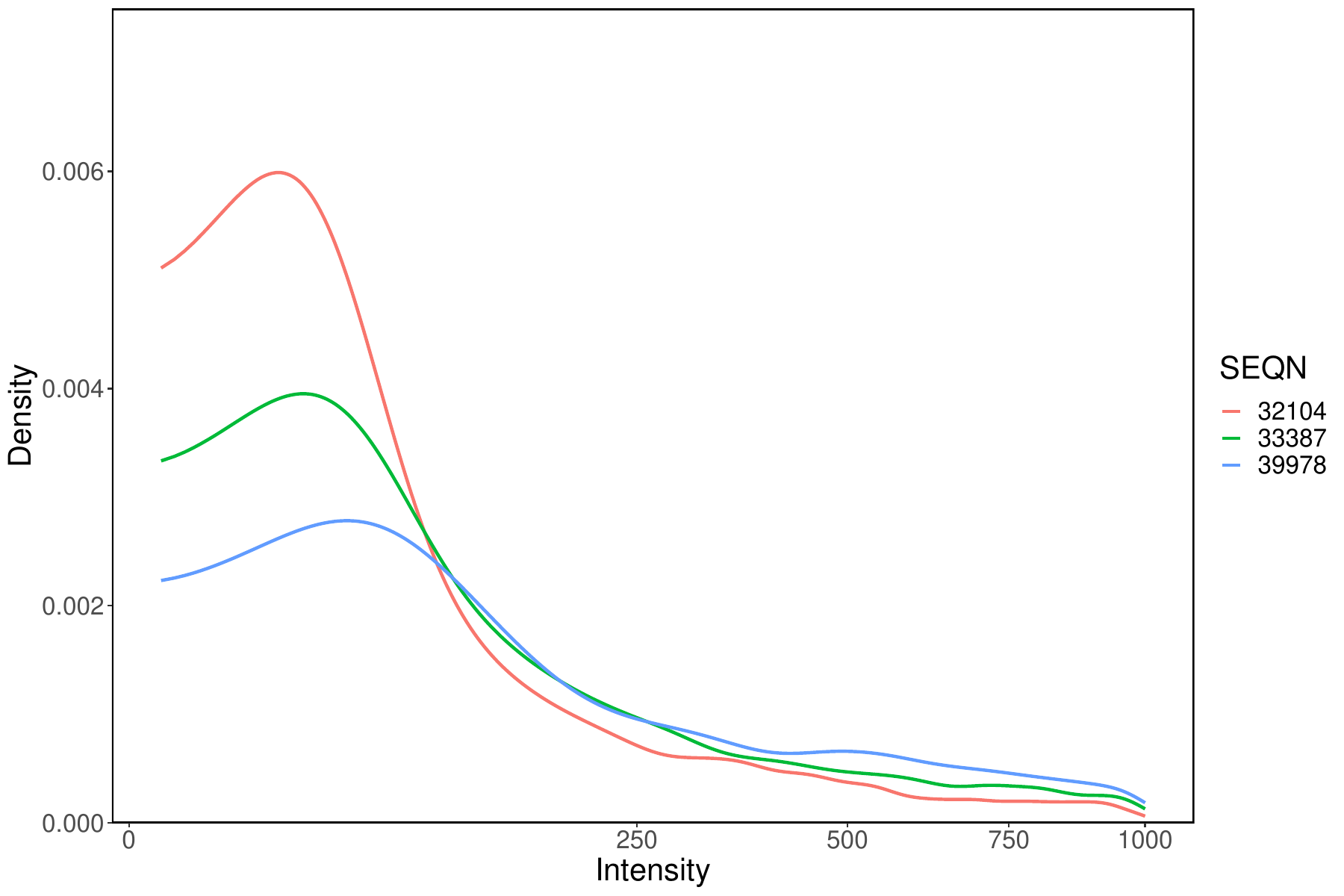}
        \caption{Density curves.}
        \label{fig:density_medoid}
      \end{subfigure}
      \caption{Left plot: the trajectory of activity values during 7 consecutive days for a participant with subject ID  31131 in the physical monitor data. Right plot: Three medoid density curves by performing a $k$-medoid cluster algorithm with $k = 3$. A square root transformation is performed on the x-axis.
      }
    \end{figure}

    We apply all five estimators described in Section \ref{sec:simulation} to a physical activity monitoring dataset. The dataset is extracted from the National Health and Nutrition Examination Survey (NHANES) 2005-2006.
    This dataset contains the activity intensity values measured by activity monitors. 
    For each participant, the physical activity intensity, ranging from 0 to 32767 cpm, was recorded every minute for 7 consecutive days. See Figure \ref{fig:traj} for an illustration.  
    More details on the physical activity measurements in this dataset can be found in 
    \url{https://wwwn.cdc.gov/Nchs/Nhanes/2005-2006/PAXRAW_D.htm}. 
    Other variables collected in this dataset are accessible from \url{https://wwwn.cdc.gov/Nchs/Nhanes/2005-2006/DEMO_D.htm}, where we extract 
    the covariates and outcomes. In this analysis, the covariates \(\bm X\) include  age, education level and family poverty income ratio, and the outcome \(Y\) is the body mass index (BMI). We aim to study the effect of activity profiles on BMI values using this dataset.

    In our analysis, we focus on white male subjects of age 20--50. We follow the similar pre-processing steps described in \cite{lin2021causal} to deal with the observations of intensity values. 
    First, we exclude observations whose reliability is questionable following NHANES protocol. 
    Then for every subject, 
    we remove the observations with intensity values higher than 1000 or equal to 0. 
    Lastly, we remove subjects whose remaining observations with intensity values are less than 100 or missing in any covariate. 
    The sample size is 427 after the pre-processing.

    The raw activity intensity profiles across different subjects are not aligned and thus generally incomparable. To address this problem, one may use their distribution functions to represent them.  
    But these intensity distributions lie in a manifold space.
    Instead,
    we apply the  kernel mean embedding \citep{Muandet-Fukumizu-Sriperumbudur17} with a Gaussian kernel to
    generate distribution representations in Hilbert space,
    as discussed in Remark \ref{rmk:kernel_example}. Hence eventually, the treatments \(A\) are taken as the kernel mean embeddings of the intensity distributions. 
    We follow the same procedures in Section \ref{sec:simulation}, including the choices of kernels and tuning parameters, to obtain all the five estimators.

    To provide a clear look at how different the estimated causal effects are provided by different estimators, 
    we present the estimated BMI values for three representative density curves chosen by performing a cluster analysis using a $k$-medoid cluster algorithm with $k = 3$ on the density curves in the dataset.
    Figure \ref{fig:density_medoid} shows three medoid density curves  while Table \ref{tab:density_medoid} shows the corresponding fitted BMI values produced by different estimators for these three curves. 
    \modify{All five estimators provide the same order of BMI values for these three curves, indicating that a more active person tends to have a lower BMI. The result of \npfcbps{} is rather similar to that of \nw{}, while  \proposed{}, \reg{} and \fcbps{} are more similar, and show a greater difference in BMI values between the curve with ID 33387 and 32104. }

\begin{figure}[h]
      \centering
      \includegraphics[width = 1\textwidth]{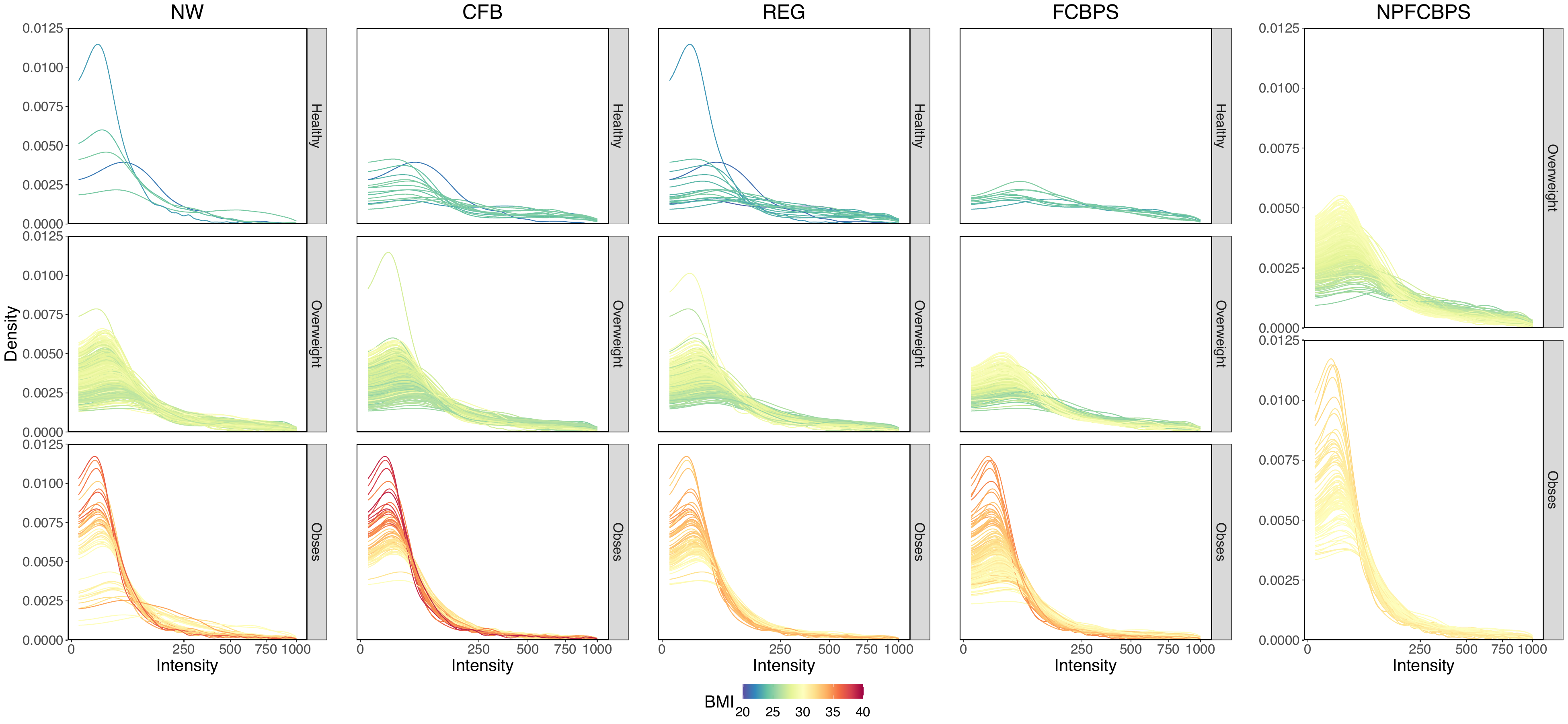}
      \caption{Density curves  divided into different categories by their corresponding estimated BMI values using different estimators (\nw{}, \proposed{}, \reg{}, \fcbps{} and \npfcbps{}). A square root transformation is performed on the x-axis.}
      \label{fig:densities}
    \end{figure}

    Next we study the performances of the five methods in categorizing the level of obesity in terms of the BMI value. According to the US Centers for Disease Control and Prevention, in terms of the BMI value, an adult 
    may be categorized as:  
    underweight (BMI \(\leq 18.5\)), healthy (\(18.5 <\) BMI \(\leq 25\)), overweight (\(25 <\) BMI \(\leq 30\)) and obese (BMI \(> 30\)). We combine the underweight and healthy categories in our dataset because it only has three underweight observations.  We visualize every unique activity profiles and their estimated BMI values (\(\hat \tau(A)\)) in Figure \ref{fig:densities}.  For a better visualization, we stratify the collection of curves by the estimated BMI categories.    
    \modify{\nw{} and \npfcbps{} categorize almost all the subjects as either overweight or obese. Moreover, the results of \nw{} are counter-intuitive since there are clearly two subgroups within the designated obese group, with one subgroup of individuals whose activities are strenuous. 
    The results of \proposed{} are similar to those of \reg{} and \fcbps{}, all indicating a steady and inverse relationship between strenuous activities and BMI.}

    \begin{table}[h]
    \caption{Estimated BMI values for the three representative density curves in Figure \ref{fig:density_medoid} by different estimators. SEQN indicates the respondent sequence number, i.e., subject ID. 
    } 
    \label{tab:density_medoid}
    \centering
    \begin{tabular}{r | rrrrr}
      \hline
     SEQN & \proposed{} & \fcbps{} & \npfcbps{} & \reg{} & \nw{} \\ 
      \hline
     33387& 27.53 & 29.01 & 28.80 & 28.48 & 28.84 \\ 
       32104 & 30.72 & 31.62 & 30.28 & 31.37 & 29.30 \\ 
     39978 & 26.11 & 27.14 & 27.59 & 26.87 & 26.60 \\ 
       \hline
    \end{tabular}
    \end{table}


\section{Discussion}
In this paper, we establish a novel covariate balancing framework for FTE estimation. 
Our framework adopts the highly flexible weight-modified kernel ridge
regression to characterize the FTE on the outcome.
The proposed weights are obtained by balancing an RKHS of the functional treatment and can be computed efficiently. 
The proposed FTE estimator is guaranteed to achieve the optimal rate of convergence without any smoothness assumptions of the oracle weight function. Its appealing empirical performance is demonstrated in an extensive simulation study and a real data application.

In the following, we outline several directions for future work. Assumption \ref{assum:ignorability} can be restrictive in practice as it requires all the covariates $\bm  X$ that adjust the dependence between $\{Y(a), a \in \mathcal{A}\}$ and $A$ are observed. 
However, this is often not guaranteed in practice, which leads to the presence of unmeasured confounding.
Inspired by the recent development in causal inference that tackles unmeasured confounding using instrumental or proxy variables, we will investigate FTE estimation
while relaxing Assumption \ref{assum:ignorability} to allow for unmeasured confounding.  In addition, while Theorem \ref{thm:emp_what_regress} provides a convergence result for $\hat \tau$ with respect to the empirical norm, 
there are fundamental difficulties in calculating the $\mathcal{L}_2$ norm for functions with functional inputs numerically, and obtaining the $\mathcal{L}_2$ norm convergence rate.  Obtaining such results probably requires a modification of our method to smooth the weights in order to establish the convergence of the estimated weight function.

\section*{Acknowledgements}
The work of Jiayi Wang is partly supported by the National Science Foundation. The work of Raymond Wong is partly  supported by the National Science Foundation. Portions of this research were conducted with the advanced computing resources provided by Texas A\&M High Performance Research Computing. The work of Xiaoke Zhang is partly supported by the George Washington University University Facilitating Fund and Columbian College of Arts and Sciences Impact Award. The work of Kwun Chuen Gary Chan is partly supported by the National Institutes of Health and National Science Foundation.

\appendix

\section{Algorithm}
\label{sec:algorithm_table}
\begin{algorithm}[H]
    \SetAlgoLined
    \caption{Outlines for obtaining $\hat \tau$. }
    \label{algo:tuning}
    \SetKwInOut{init}{Initialization}
    \KwIn{Observed confounders \(\bm  X_i \in \mathbb{R}^p\) and approximated treatments \(\hat A_i \in \mathcal{A}\), \(i = 1,\dots, n\); smoothing parameter \(\lambda>0\); the fitted outcome regression model $\hat m$ and a sequence of tuning parameters \(\eta_k\), \(k = 1,\dots, K\). 
    }
    Calculate \(\bm G_A\), \(\bm G_X\), \(\bar{\bm G}\) and \(\bar{g}_X\) according to \eqref{eqn:gram_matrices}.\\
    Decompose \(\bm G_F\) and obtain \(\bm M_1\), \(\bm M_2\) and \(\bm M\).
    \BlankLine
    \For {$k = 1,\dots, K$}{
        Optimize \eqref{eqn:estw_opt} by L-BFGS-B algorithm with  \(\eta = \eta_{k}\) and obtain solution \(\hat{\bm w}^{(\eta_k)}\).\\
        Compute the value of $V(\eta_k)$ from \eqref{eqn:eta_tune}.
     
    }
    Select  \(\tilde \eta\) such that $V(\tilde \eta)$ is the smallest among all $V(\eta_k)$, $k = 1,\dots, K$. 
    
    Construct the adjusted response \(Z_i = \hat{ w}^{(\tilde \eta)}_i Y_i\), \(i = 1,\dots,n\).

    Fix the  response as  \(Z_i\), \(i = 1,\dots,n\) and obtain 
    $$\hat \tau = \left( \frac{1}{n}\sum_{i=1}^n \mathcal{S}_{\hat A_i} + \lambda \mathcal{I}\right)^{-1}\left( \frac{1}{n}\sum_{i=1}^n \mathcal{K}_{\hat A_i} Z_i\right).$$
    
    \KwOut{$ \hat {\bm w}^{\tilde \eta}$ and $\hat \tau$. }

  \end{algorithm}

  \section{Proof}
  \begin{proof}[\bf Proof of Proposition \ref{prop:tau}]    
      Given the discussion of Assumptions in Section \ref{sec:theory}, under Assumption \ref{assum:kernelbound}, the spectral theorem gives 
      \begin{align*}
          \mathcal{S} = \sum_{l=1}^L t_l \langle \cdot, e_l \rangle_{\Hscr_A} e_l,
      \end{align*}
      where $0 < t_{l+1} \leq t_l$,  $\{e_l\}_{l=1}^\infty$ is a basis of $\mathrm{Ker}\mathcal{\mathcal{S}}^{\perp}$. Define the operator $T:\mathcal{L}_2(\rho_A) \rightarrow \mathcal{L}_2(\rho_A)$ to be the integral operator of kernel $K_A$,
      \begin{align*}
          (T\rho)(a) = \E K_A(a, A )\psi(A) = \int_{\mathcal{A}}  K_A(a, a') \psi(a') d\rho_A(a')
      \end{align*}
      for $\psi \in \mathcal{L}_2(\rho_A)$.
      Based on Remark 2 in \cite{caponnetto2007optimal}, it can be shown that 
     \begin{align*}
         T = \sum_{l=1}^K \nu^A_l \langle \cdot, \phi^A_l \rangle_{\rho_A} \phi_l^A,
     \end{align*}
     where $\{\phi^A_l\}_{l=1}^L$ is basis of $\mathrm{Ker}T^\perp$, $T^{1/2} \phi^A_l = t_l^{1/2} \phi^A_l = e_l$ and $\lambda_A^l = t_l$, $l=1,\dots, L$. And we have 
     \begin{align*}
         K_A(a_1, a_2) = \sum_{l=1}^L \nu^A_l \phi_l^A(a_1) \phi^A_l(a_2).
     \end{align*}
     As $K_X$ is a bounded continuous positive definite kernel,  Mercer Theorem yields 
     \begin{align*}
         K_X(\bm x_1,\bm x_2) = \sum_{l=1}^L \nu^X_l \phi^X_l(\bm x_1) \phi^X_l(\bm x_2),
     \end{align*}
     where $0<\phi_{l+1}^X \leq \phi^X_{l}$ are eigenvalues, $\{(\phi^X_l\}_{l=1}^\infty$  is a set of  orthonormal basis in $\mathcal{L}_2(\rho_X)$ and $\{(\nu^X_l)^{1/2}\phi^X_l\}_{l=1}^\infty$ is a set of  orthonormal basis in $\Hscr_X$. 
  
     Next, we prove the direction that $\E_{\bm X \sim \rho_X} g(\cdot, \bm X) \in \Hscr_A$ for any $g \in \Hscr$.
     
     Given these two sets of basis, for any function $g \in \Hscr$, it can be expressed as 
     \begin{align*}
         g(a,\bm x) = \sum_{l_1, l_2} \beta_{l_1, l_2} \sqrt{\nu_{l_1}^A} \sqrt{\nu_{l_2}^X}\phi^A_l(a) \phi^X_l(\bm x),
     \end{align*}
     for some coefficients $\{\beta_{l_1,l_2}\}_{l_1,l_2=1}^{\infty}$ with 
     \begin{align*}
         \|g\|^2_{\Hscr} = \sum_{l_1,l_2} \beta_{l_1, l_2}^2  < \infty.
     \end{align*}
     Then we have 
     \begin{align*}
         \|\E_{\bm X \sim \rho_X} g(\cdot, \bm X) \|^2_{\Hscr_A} = \sum_{l_1} (\beta_{l_1,l_2})^2 \left\{\E \left(  \sum_{l_2} \sqrt{\nu_{l_2}^X}\phi_{l_2}^X(\bm X)  \right)\right\}^2 
          \leq \sum_{l_1} (\beta_{l_1,l_2})^2\E \left(  \sum_{l_2} \sqrt{\nu_{l_2}^X} \phi_{l_2}^X(\bm X)  \right)^2  \\
          \leq \sum_{l_1} (\beta_{l_1,l_2})^2 \sum_{l_2}\nu_{l_2}^X (\beta_{l_2}^X)^2
            \leq \left( \max_{l_2} \nu_{l_2}^X \right) \sum_{l_1}\sum_{l_2} (\beta^A_{l_1,l_2})^2 = \nu_{1}^X (\sum_{l_1,l_2} \beta_{l_1, l_2}^2) = \nu_{1}^X \|g\|^2_{\Hscr},
     \end{align*}
     the second inequality is due to the fact that $\{\phi^X_l\}_{l=1}^\infty$ is orthonormal in $\mathcal{L}_2(\rho_X)$.
     Note that $K_X$ is bounded, thus $\nu_1^X$ is  bounded. The conclusion follows.
  
     Now, we prove the direction that for any function $f\in \Hscr_A$, there exists a function $g \in \Hscr$, such that $\E_{\bm X\sim \rho_X} g(\cdot, \bm X)\in \Hscr$.
  
  First, note that these exists a function $u\in \Hscr_X$, such that 
  \begin{align*}
      \E u(\bm X) = \int_{\bm x} u(\bm x) \rho_X(\bm x) d \bm x \neq 0.
  \end{align*}
  
  Take 
  \begin{align*}
      g(a , \bm x ) = f(a) \frac{u(\bm x)}{ \E u(\bm X)}.
  \end{align*}
  One can verify that $   \E_{\bm X \sim \rho_X} g(\cdot, \bm X) = f(\cdot)$. Since
  $\Hscr = \Hscr_A \otimes \Hscr_X$, $f \in \Hscr_A$ and ${u(\bm x)}/{ \E u(\bm X)} \in \Hscr_X$, therefore $g \in \Hscr$.
  Therefore, the conclusion is verified.
  \end{proof}

  \begin{proof}[\bf Proof of Theorem \ref{thm:representer}]
  \label{proof:representer}
      Take $\Hscr_n^{\perp}$ as the orthogonal space of $\Hscr_n$. For any function $u \in \Hscr$, we can decompose it into two orthogonal parts $u_1$ and $u_2$ such that $u_1 \in \Hscr_n$ and $u_2 \in \Hscr_n^{\perp}$.  Then for any $u \in \Hscr$, 
      \begin{align*}
         &  Q(\bm w, \lambda, u)\\
         & = \frac{1}{n}\sum_{k=1}^n \left[ \left\{ \left( \frac{1}{n}\sum_{i=1}^n \mathcal{S}_{\hat A_i} + \lambda  \mathcal{I} \right)^{-1}\left( \frac{1}{n}\sum_{i=1}^n \mathcal{K}_{\hat A_i} w_i u(\hat A_i, \bm X_i)\right) \right\}(A_k)   - \frac{1}{n}\sum_{j=1}^n u(A_k,\bm  X_j)  \right]^2 \\
           & =  \frac{1}{n}\sum_{k=1}^n \left[ \left\{ \left( \frac{1}{n}\sum_{i=1}^n \mathcal{S}_{\hat A_i} + \lambda  \mathcal{I}\right)^{-1}\left( \frac{1}{n}\sum_{i=1}^n \mathcal{K}_{\hat A_i} w_i \langle K((\hat A_i,\bm  X_i), (\cdot, \cdot)) , u \rangle \right) \right\}(A_k)  \right. \\
         &  \left. - \frac{1}{n}\sum_{j=1}^n  \langle K(\hat A_k,\bm  X_j), (\cdot, \cdot)) , u \rangle \right]^2\\
         =  &\frac{1}{n}\sum_{k=1}^n \left[ \left\{ \left( \frac{1}{n}\sum_{i=1}^n \mathcal{S}_{\hat A_i} + \lambda  \mathcal{I}\right)^{-1}\left( \frac{1}{n}\sum_{i=1}^n \mathcal{K}_{\hat A_i} w_i \langle \kernel_A(\cdot, \hat A_i) \kernel_X(\cdot, \bm X_i), (\cdot, \cdot)) , u_1 + u_2 \rangle \right) \right\}(A_k)  \right. \\
         &  \left. -   \left \langle \kernel_A(\cdot, \hat A_k) \left\{\frac{1}{n}\sum_{j=1}^n \kernel_X(\cdot, \bm  X_j)\right\}, (\cdot, \cdot)) , u_1 + u_2 \right \rangle \right]^2 \\
         & =  \frac{1}{n}\sum_{k=1}^n \left[ \left\{ \left( \frac{1}{n}\sum_{i=1}^n \mathcal{S}_{\hat A_i} + \lambda  \mathcal{I}\right)^{-1}\left( \frac{1}{n}\sum_{i=1}^n \mathcal{K}_{\hat A_i} w_i \langle \kernel_A(\cdot, \hat A_i) \kernel_X(\cdot,\bm   X_i), (\cdot, \cdot)) , u_1  \rangle \right) \right\}(A_k)  \right. \\
         &  \left. -   \left \langle \kernel_A(\cdot, \hat A_k) \left\{\frac{1}{n}\sum_{j=1}^n \kernel_X(\cdot, \bm  X_j)\right\}, (\cdot, \cdot)) , u_1 \right \rangle \right]^2  =  Q(\bm w, \lambda, u_1). 
      \end{align*}
      On the other hand,  $\| u\|_\Hscr = \| u_1 \|_\Hscr + \| u_2 \|_\Hscr $. For any $u$ such that $\|u\|_\Hscr = 1$ and $\|u_1\|_{\Hscr} > 0 $, we can always find another $\tilde{u} = \frac{\|u\|_\Hscr}{\|u_1\|_\Hscr} u_1 \in \Hscr_n $.  It is easy to verify that $\|\tilde{u}\|_{\Hscr} = 1$ and 
      \begin{align*}
           Q(\bm w, \lambda, \tilde{u}) = & Q\left(\bm w, \lambda, \frac{\|u\|_\Hscr}{\|u_1\|_\Hscr} u_1\right) 
           = \left(\frac{\|u\|_\Hscr}{\|u_1\|_\Hscr}\right)^2 Q\left(\bm w, \lambda, u_1\right)  
           =  \left(\frac{\|u\|_\Hscr}{\|u_1\|_\Hscr}\right)^2 Q\left(\bm w, \lambda, u\right) 
           \ge  Q\left(\bm w, \lambda, u\right).  
      \end{align*}
      The conclusion follows. 
  \end{proof}

  \begin{proof}[\bf Proof of Lemma \ref{lem:convexity}]
  \label{proof:convexity}
      Consider any vectors $\bm w_1 \in \mathbb{R}^{n}$ and $\bm w_2 \in \mathbb{R}^{n}$, and $t \in [0,1]$. For $\bm \beta \in \mathbb{R}^q$, we have 
      \begin{align*}
     & \left\| \left[ \bm A \mathrm{diag}\{ t\bm w_1 + (1-t) \bm w_2\} \bm B -  \bm D \right] \bm \beta \right\|_2^2 \\
     = & \left\| t \left[ \bm A \mathrm{diag}(\bm w_1) \bm B -  \bm D \right]\bm \beta +  (1-t) \left[ \bm A \mathrm{diag}(\bm w_2) \bm B -  \bm D \right]\bm \beta \right\|_2^2 \\
     \leq & t \left\| \left[ \bm A \mathrm{diag}(\bm w_1) \bm B -  \bm D \right]\bm \beta \right\|_2^2 +  (1-t) \left\| \left[ \bm A \mathrm{diag}(\bm w_2) \bm B -  \bm D \right]\bm \beta \right\|_2^2.
      \end{align*}
      The inequality is due to the convexity of $\|\cdot\|_2^2$.  
      Suppose that $\bm \beta$ is the right singular vector of $\bm A \mathrm{diag}\{ t\bm w_1 + (1-t) \bm w_2\} \bm B -  \bm D$ that corresponds to the largest singular value.  Then 
  \begin{align*}
      &\left[ \sigma_{\max} \left\{ \bm A \mathrm{diag}(\bm w)\bm B -  \bm D \right\} \right]^2\\
      &= \left\| \left[ \bm A \mathrm{diag}\{ t\bm w_1 + (1-t) \bm w_2\} \bm B -  \bm D \right] \bm \beta \right\|_2^2 \\
      & \leq  t \left\| \left[ \bm A \mathrm{diag}(\bm w_1) \bm B -  \bm D \right]\bm \beta \right\|_2^2 +  (1-t) \left\| \left[ \bm A \mathrm{diag}(\bm w_2) \bm B -  \bm D \right]\bm \beta \right\|_2^2 \\
      & \leq t \left[ \sigma_{\max} \left\{ \bm A \mathrm{diag}(\bm w)\bm B -  \bm D \right\} \right]^2 + (1-t) \left[ \sigma_{\max} \left\{ \bm A \mathrm{diag}(\bm w_2)\bm B -  \bm D \right\} \right]^2.
  \end{align*}
  The second inequality is due to the definition of the largest singular value. Then conclusion follows.
  \end{proof}

      \begin{proof}[\bf Proof of Theorem \ref{thm:emp_bal_error_truew}]
  
  First, we derive the bounds for $\sup_{u \in \Hscr(1)}Q(\bm \truew, \lambda,u )$ and $R(\bm \truew, \lambda)$
      
        Take the function \(z(a;u) = \E_{\bm  X\sim\rho_X}u(a, X)\) for \(u \in \Hscr(1)\).  Take \(\mathcal{S}_{A} = \frac{1}{n}\sum_{i=1}^n \mathcal{S}_{A_i} \), \(\mathcal{S}_{\hat A} = \frac{1}{n}\sum_{i=1}^n \mathcal{S}_{\hat A_i} \).
        It's easy to see that \(\sup_{u\in \Hscr(1)}Q(w, \lambda,u)\) can be bounded by the following components:
        \begin{align}
          & \sup_{u\in \Hscr(1)} Q( \bm \truew, \lambda, u) \nonumber \\
          &  \lesssim \sup_{u\in \Hscr(1)} \left\| \left( \frac{1}{n}\sum_{i=1}^n \mathcal{S}_{\hat A_i} + \lambda  \mathcal{I}\right)^{-1}\left[\frac{1}{n}\sum_{i=1}^n \left( \mathcal{K}_{\hat A_i}\left\{\truew_i  u(\hat A_i, \bm  X_i) - z(\hat A_i;u)\right\} \right. \right. \right. \nonumber \\
          & \qquad \left. \left.  \left.  - \mathcal{K}_{A_i} \left\{\truew_i  u(A_i, \bm  X_i) - z(A_i;u)\right\}\right) \right]  \right\|_{n}^2  \label{eqn:comp-1}\\
          & +  \sup_{u\in \Hscr(1)} \left\| \left( \mathcal{S}_{\hat A} + \lambda  \mathcal{I}\right)^{-1} \left( \frac{1}{n}\sum_{i=1}^n \mathcal{S}_{ A_i} - \frac{1}{n}\sum_{i=1}^n \mathcal{S}_{\hat A_i}\right)\left(  \mathcal{S}_{A} + \lambda  \mathcal{I}\right)^{-1}  \right. \nonumber \\
          & \qquad \left. \left[\frac{1}{n}\sum_{i=1}^n \mathcal{K}_{A_i} \left\{\truew_i  u(A_i, \bm  X_i) - z(A_i;u)\right\}\right]  \right\|_{n}^2 \label{eqn:comp0}\\
          &  + \sup_{u\in \Hscr(1)} \left\| \left( \frac{1}{n}\sum_{i=1}^n \mathcal{S}_{A_i} + \lambda  \mathcal{I} \right)^{-1}\left[\frac{1}{n}\sum_{i=1}^n \mathcal{K}_{A_i} \left\{\truew_i  u(A_i, \bm  X_i) - z(A_i;u)\right\}\right]  \right\|_{n}^2 \label{eqn:comp1}\\
           & + \sup_{u\in \Hscr(1)} \left\| \left( \frac{1}{n}\sum_{i=1}^n \mathcal{S}_{A_i} + \lambda  \mathcal{I}\right)^{-1}\left[\frac{1}{n}\sum_{i=1}^n \mathcal{K}_{A_i}  z(A_i;u)\right] - z(\cdot;u) \right\|_{n}^2 \label{eqn:comp2}\\
            & + \sup_{u\in \Hscr(1)} \left\| z(\cdot;u) -  \frac{1}{n}\sum_{j=1}^n u(\cdot, \bm 
   X_j)\right\|_{n}^2 \label{eqn:comp3}
        \end{align}
        Next, we consider to bound \eqref{eqn:comp-1}, \eqref{eqn:comp0}, \eqref{eqn:comp1}, \eqref{eqn:comp2} and \eqref{eqn:comp3} one by one. Note that under the fully observed case (Assumption \ref{assume:function_approximation} 1), $\hat A_i = A_i$ for $i =1,\dots, n$, and we only need to focus on \eqref{eqn:comp1} - \eqref{eqn:comp3}.

    First, follow the proof in  Section A.1.11 of \cite{szabo2015two}, by Assumption \ref{assume:function_approximation}, we have
    \begin{align*}
      \left \|  \left( \frac{1}{n}\sum_{i=1}^n \mathcal{S}_{\hat A_i}  \right) -  \left( \frac{1}{n}\sum_{i=1}^n \mathcal{S}_{ A_i} \right) \right \|^2_{\mathcal{L}_{\Hscr_A}} \leq \frac{1}{n} \sum_{i=1}^n  \left \| \mathcal{S}_{\hat A_i}  -   \mathcal{S}_{ A_i}  \right \|^2_{\mathcal{L}_{\Hscr_A}} =\bigOp\left( {H^2 \oldu{cnt:kernelbound}^{h} \kappa^{2h}} \right).  
    \end{align*}

  And note that for any function \(f \in \Hscr_A\), 
  \begin{align*}
    \|f\|^2_{n} = \left\|\left(\frac{1}{n}\sum_{i=1}^n \mathcal{S}_{\hat A_i}  \right)^{1/2} f\right\|^2_{\Hscr_A}.
  \end{align*}

        \begin{itemize}
  
  \item For \eqref{eqn:comp-1}. 
   \begin{align*}
     &\eqref{eqn:comp-1}\\
     &\leq \left\| \left( \frac{1}{n}\sum_{i=1}^n \mathcal{S}_{\hat A_i} \right)^{1/2}  \left( \frac{1}{n}\sum_{i=1}^n \mathcal{S}_{\hat A_i} + \lambda  \mathcal{I} \right)^{-1} \right\|^2_{\mathcal{L}_{\Hscr_A}} \\
    &\qquad \times  \left\|\frac{1}{n}\sum_{i=1}^n \left( \mathcal{K}_{\hat A_i}\left\{\truew_i  u(\hat A_i, \bm X_i) - z(\hat A_i;u)\right\}  - \mathcal{K}_{A_i} \left\{\truew_i  u(A_i, \bm X_i) - z(A_i;u)\right\}\right)   \right\|_{\Hscr_A}^2.
   \end{align*}
   By the spectral theorem,
   \begin{align*}
    \left\| \left( \frac{1}{n}\sum_{i=1}^n \mathcal{S}_{\hat A_i} \right)^{1/2}  \left( \frac{1}{n}\sum_{i=1}^n \mathcal{S}_{\hat A_i} + \lambda  \mathcal{I}\right)^{-1} \right\|^2_{\mathcal{L}_{\Hscr_A}} \leq \frac{1}{\lambda}.
   \end{align*}
  
   And by H\"{o}lder continuous in Assumption \ref{assume:function_approximation}, 
   \begin{align*}
   & \left\|\frac{1}{n}\sum_{i=1}^n \left( \mathcal{K}_{\hat A_i}\left\{\truew_i  u(\hat A_i, \bm 
   X_i) - z(\hat A_i;u)\right\}  - \mathcal{K}_{A_i} \left\{\truew_i  u(A_i, \bm  X_i) - z(A_i;u)\right\}\right)  \right\|_{\Hscr_A}^2 \\
     \lesssim  & \frac{H^2}{n}\sum_{i=1}^n \left[ d(A_i, \hat A_i)^{2h} \oldu{cnt:overlapupper}^2  \oldu{cnt:kernelbound}^4 \right]
     =  \bigOp \left( {H^2 \oldu{cnt:overlapupper}^2 \oldu{cnt:kernelbound}^4 } \kappa^{2h} \right).
   \end{align*}
  
   Then 
   \begin{align*}
    \eqref{eqn:comp-1}  = \bigOp \left( \frac{H^2 \oldu{cnt:overlapupper}^2 \oldu{cnt:kernelbound}^4 \kappa^{2h}}{\lambda} \right).
   \end{align*}

  \item For \eqref{eqn:comp0},  by similar proof, we can show that 
  \begin{align*}
    &\eqref{eqn:comp0}\\
    &\leq \left\|  \left( \frac{1}{n}\sum_{i=1}^n \mathcal{S}_{\hat A_i} \right)^{1/2} \left( \frac{1}{n}\sum_{i=1}^n \mathcal{S}_{\hat A_i} + \lambda  \mathcal{I}\right)^{-1} \right\|^2_{\mathcal{L}_{\Hscr_A}}  \left\| \left( \frac{1}{n}\sum_{i=1}^n \mathcal{S}_{ A_i} - \frac{1}{n}\sum_{i=1}^n \mathcal{S}_{\hat A_i}\right)\right\|^2_{\mathcal{L}_{\Hscr_A}}  \\
    & \qquad \times  \sup_{u\in \Hscr(1)} \left\| \left(  \mathcal{S}_{A} + \lambda  \mathcal{I}\right)^{-1} \left[\frac{1}{n}\sum_{i=1}^n \mathcal{K}_{A_i}  \left\{\truew_i  u(A_i, \bm X_i) - z(A_i;u)\right\}\right]  \right\|_{\Hscr_A}^2 \\
    &=  \bigOp \left( \frac{H^2 \oldu{cnt:overlapupper}^2 \oldu{cnt:kernelbound}^4  \oldu{cnt:kernelbound}^h \kappa^{2h}}{\lambda} \right) \frac{1}{\lambda}\\
    &\qquad \times \sup_{u\in \Hscr(1)} \left\| \left(  \mathcal{S}_{A} + \lambda  \mathcal{I}\right)^{-\frac{1}{2}} \left[\frac{1}{n}\sum_{i=1}^n \mathcal{K}_{A_i}  \left\{\truew_i  u(A_i, \bm X_i) - z(A_i;u)\right\}\right]  \right\|_{\Hscr_A}^2 .
  \end{align*}
  By later argument (see the proof for bounding \eqref{eqn:comp1}), we can prove that 
  \[\sup_{u\in \Hscr(1)} \left\| \left(  \mathcal{S}_{A} + \lambda  \mathcal{I}\right)^{-\frac{1}{2}} \left[\frac{1}{n}\sum_{i=1}^n \mathcal{K}_{A_i}  \left\{\truew_i  u(A_i, \bm X_i) - z(A_i;u)\right\}\right]  \right\|_{\Hscr_A}^2 = \bigOp\left( \frac{\mathcal{N}(\lambda)}{n} \right).\]
  Then by the condition $\kappa^{2h} = \bigO(\lambda^2)$, we have
  \begin{align*}
    \eqref{eqn:comp0} = \bigOp\left( \frac{H^2 \oldu{cnt:overlapupper}^2 \oldu{cnt:kernelbound}^4  \oldu{cnt:kernelbound}^h \kappa^{2h}}{\lambda} \frac{\mathcal{N}(\lambda)}{n \lambda} \right) = \bigOp\left( \frac{\mathcal{N}(\lambda)}{n} \right).
  \end{align*}
  
            \item     Next, we focus on controlling term \eqref{eqn:comp1}.
          
      We have 
    
        \begin{align}
            & \sup_{u\in \Hscr(1)} \left\| \left( \frac{1}{n}\sum_{i=1}^n \mathcal{S}_{A_i} + \lambda  \mathcal{I} \right)^{-1}\left[\frac{1}{n}\sum_{i=1}^n \mathcal{K}_{A_i} \left\{\truew_i  u(A_i, \bm X_i) - z(A_i;u)\right\}\right] \right\|_{n}^2 \nonumber\\
           &= \sup_{u\in \Hscr(1)} \left\|\left(\frac{1}{n}\sum_{i=1}^n \mathcal{S}_{ \hat A_i}  \right)^{1/2}  \left( \frac{1}{n}\sum_{i=1}^n \mathcal{S}_{A_i} + \lambda  \mathcal{I}\right)^{-1}\left[\frac{1}{n}\sum_{i=1}^n \mathcal{K}_{A_i} \left\{\truew_i  u(A_i, \bm X_i) - z(A_i;u)\right\}\right]\right\|^2_{\Hscr_A} \nonumber\\
           & \leq  \left\|\left(\frac{1}{n}\sum_{i=1}^n \mathcal{S}_{\hat A_i}  \right)^{1/2}  \left( \frac{1}{n}\sum_{i=1}^n \mathcal{S}_{A_i} + \lambda  \mathcal{I}\right)^{-1} (\mathcal{S} + \lambda)^{\frac{1}{2}} \right\|^2_{\mathcal{L} (\Hscr_A)} \nonumber\\
           & \qquad \times \sup_{u\in \Hscr(1)} \left\| (\mathcal{S} + \lambda)^{-\frac{1}{2}} \left[\frac{1}{n}\sum_{i=1}^n \mathcal{K}_{A_i} \left\{\truew_i  u(A_i, \bm X_i) - z(A_i;u)\right\}\right]\right\|^2_{\Hscr_A}. \label{eqn:first_term}
           \end{align}

           We start with bounding  the first term:
        \begin{align*}
            &\left\|\left(\frac{1}{n}\sum_{i=1}^n \mathcal{S}_{\hat A_i}  \right)^{\frac{1}{2}} (S_{\bm A} + \lambda  \mathcal{I} )^{-1}  (S + \lambda  \mathcal{I})^{\frac{1}{2}}\right\|_{\mathcal{L}(\Hscr_A)}\\
             = & \left\| \left(\frac{1}{n}\sum_{i=1}^n \mathcal{S}_{\hat A_i}  \right)^{\frac{1}{2}} \left( \mathcal{S} + \lambda  \mathcal{I}\right)^{-\frac{1}{2}} \left\{ I - \left( \mathcal{S}+ \lambda  \mathcal{I}\right)^{-\frac{1}{2}} \left( \mathcal{S} -  \left(\frac{1}{n}\sum_{i=1}^n \mathcal{S}_{A_i}  \right) \right)\left( \mathcal{S} + \lambda  \mathcal{I}\right)^{\frac{1}{2}}\right\}^{-1}\right\|_{\mathcal{L}(\Hscr_A)}\\
             \leq & \left\|  \left(\frac{1}{n}\sum_{i=1}^n \mathcal{S}_{\hat A_i}  \right)^{\frac{1}{2}} \left( \mathcal{S} + \lambda  \mathcal{I}\right)^{-\frac{1}{2}}\right\|_{\mathcal{L}(\Hscr_A)} \nonumber \\ 
            & \qquad \times \left\|  \left\{ I - \left( \mathcal{S}+ \lambda  \mathcal{I}\right)^{-\frac{1}{2}} \left( \mathcal{S} -  \left(\frac{1}{n}\sum_{i=1}^n \mathcal{S}_{A_i}  \right) \right)\left( \mathcal{S} + \lambda  \mathcal{I}\right)^{-\frac{1}{2}}\right\}^{-1}\right\|_{\mathcal{L}(\Hscr_A)}.
           \end{align*}
           From \cite{caponnetto2007optimal}, we can show that if \(n \ge 2C_{\eta} \kappa \mathcal{N}(\lambda)/\lambda\), where \(C_\eta = 32\log^2(6/\eta)\), and \(\lambda \leq \|\mathcal{S}\|_{\mathcal{L}(\Hscr_A)}\), 
           \begin{align}
             \label{eqn:operator_diff}
             \left\| \left( \mathcal{S}+ \lambda  \mathcal{I}\right)^{-\frac{1}{2}} \left( \mathcal{S} -  \left(\frac{1}{n}\sum_{i=1}^n \mathcal{S}_{A_i}  \right) \right)\left( \mathcal{S} + \lambda  \mathcal{I}\right)^{-\frac{1}{2}}\right\|_{\mathcal{L}(\Hscr_A)} \leq \frac{1}{2},
           \end{align}
           with probability at least \(1-2\eta/3\). And therefore under the same condition, we have
           \begin{align}
             \left\|   \left\{ I - \left( \mathcal{S}+ \lambda  \mathcal{I}\right)^{-\frac{1}{2}} \left( \mathcal{S} -  \left(\frac{1}{n}\sum_{i=1}^n \mathcal{S}_{A_i}  \right) \right)\left( \mathcal{S} + \lambda  \mathcal{I}\right)^{-\frac{1}{2}}\right\}^{-1}\right\|_{\mathcal{L}(\Hscr_A)} \leq 2. 
           \end{align}
           with probability at least \(1-2\eta/3\). 
           
           Then we bound \(\|  (\sum_{i=1}^n \mathcal{S}_{\hat A_i}  /n)^{1/2} \left( \mathcal{S} + \lambda  \mathcal{I}\right)^{-1/2}\|_{\mathcal{L}(\Hscr_A)}\). Notice that 
           \begin{align*}
             \left\|  \left(\frac{1}{n}\sum_{i=1}^n \mathcal{S}_{\hat A_i}  \right)^{\frac{1}{2}} \left( \mathcal{S} + \lambda  \mathcal{I}\right)^{-\frac{1}{2}}\right\|^2_{\mathcal{L}(\Hscr_A)} = \left\|  \left( \mathcal{S} + \lambda  \mathcal{I}\right)^{-\frac{1}{2}} \left(\frac{1}{n}\sum_{i=1}^n \mathcal{S}_{\hat A_i}  \right) \left( \mathcal{S} + \lambda  \mathcal{I}\right)^{-\frac{1}{2}}\right\|_{\mathcal{L}(\Hscr_A)} \\
             \leq \left\|  \left( \mathcal{S} + \lambda  \mathcal{I}\right)^{-\frac{1}{2}} \mathcal{S} \left( \mathcal{S} + \lambda \mathcal{I} \right)^{-\frac{1}{2}}\right\|_{\mathcal{L}(\Hscr_A)} + \left\|  \left( \mathcal{S} + \lambda   \mathcal{I} \right)^{-\frac{1}{2}} \left( \mathcal{S} - \frac{1}{n}\sum_{i=1}^n \mathcal{S}_{\hat A_i}  \right) \left( \mathcal{S} + \lambda  \mathcal{I}\right)^{-\frac{1}{2}}\right\|_{\mathcal{L}(\Hscr_A)}\\
            = \left\|  \mathcal{S}^{\frac{1}{2}} \left( \mathcal{S} + \lambda   \mathcal{I} \right)^{-\frac{1}{2}}\right\|^2_{\mathcal{L}(\Hscr_A)} +  \left\|  \left( \mathcal{S} + \lambda  \mathcal{I}\right)^{-\frac{1}{2}} \left( \mathcal{S} - \frac{1}{n}\sum_{i=1}^n \mathcal{S}_{\hat A_i}  \right) \left( \mathcal{S} + \lambda  \mathcal{I}\right)^{-\frac{1}{2}}\right\|_{\mathcal{L}(\Hscr_A)}
           \end{align*}
           For the first component, it's easy to see that
           \begin{align*}
             \left\|  \left( \mathcal{S} + \lambda  \mathcal{I} \right)^{-\frac{1}{2}} \mathcal{S} \left( \mathcal{S} + \lambda  \mathcal{I}\right)^{-\frac{1}{2}}\right\|_{\mathcal{L}(\Hscr_A)} = \left\|  \mathcal{S}^{\frac{1}{2}} \left( \mathcal{S} + \lambda  \mathcal{I}\right)^{-\frac{1}{2}}\right\|^2_{\mathcal{L}(\Hscr_A)} \leq 1
           \end{align*}
           because of the spectral theorem. 
           
           And by \eqref{eqn:operator_diff}, we have
           \begin{align*}
           &   \left\|  \left( \mathcal{S} + \lambda  \mathcal{I}\right)^{-\frac{1}{2}} \left( \mathcal{S} - \frac{1}{n}\sum_{i=1}^n \mathcal{S}_{\hat A_i}  \right) \left( \mathcal{S} + \lambda  \mathcal{I}\right)^{-\frac{1}{2}}\right\|_{\mathcal{L}(\Hscr_A)} \\
            \leq &  \left\|  \left( \mathcal{S} + \lambda  \mathcal{I}\right)^{-\frac{1}{2}} \left( \mathcal{S} - \frac{1}{n}\sum_{i=1}^n \mathcal{S}_{ A_i}  \right) \left( \mathcal{S} + \lambda \mathcal{I} \right)^{-\frac{1}{2}}\right\|_{\mathcal{L}(\Hscr_A)}  \\
            & \qquad + \left\|  \left( \mathcal{S} + \lambda  \mathcal{I}\right)^{-\frac{1}{2}} \left( \frac{1}{n}\sum_{i=1}^n \mathcal{S}_{\hat  A_i}  - \frac{1}{n}\sum_{i=1}^n \mathcal{S}_{ A_i}  \right) \left( \mathcal{S} + \lambda  \mathcal{I}\right)^{-\frac{1}{2}}\right\|_{\mathcal{L}(\Hscr_A)} \\
            \leq & \bigOp(1) + \frac{1}{\lambda} \bigOp\left( {H \oldu{cnt:kernelbound}^{h/2} \kappa^{h}}\right) 
            \leq  \bigOp(1).
           \end{align*}
           The last inequality is due to the condition for $\kappa$ that $\kappa^h = \bigO(\lambda)$.

          Then we have 
           \begin{align*}
             \left\|  \left(\frac{1}{n}\sum_{i=1}^n \mathcal{S}_{\hat A_i}  \right)^{\frac{1}{2}} \left( \mathcal{S} + \lambda  \mathcal{I} \right)^{-\frac{1}{2}}\right\|^2_{\mathcal{L}(\Hscr_A)} = \bigOp\left( 1 \right).
           \end{align*}
           And overall we show that 
           \begin{align}
               \left\|\left(\frac{1}{n}\sum_{i=1}^n \mathcal{S}_{\hat A_i}  \right)^{\frac{1}{2}} (S_{\bm A} + \lambda  \mathcal{I})^{-1}  (S + \lambda \mathcal{I})^{\frac{1}{2}}\right\|_{\mathcal{L}(\Hscr_A)} = \bigOp\left( 1 \right). \label{eqn:opnorm_bound}
           \end{align}
        
        Next, we bound the second term in \eqref{eqn:first_term}.
            Take \(r_i\) as independent Rademacher random variables. It's easy to see that \(\E\{\truew_i  u(A_i, X_i) - z(A_i;u)\} = 0 \) for every \(i\) and \(u \in \Hscr(1)\).
            Due to symmetrization inequality, we have
            \begin{align}
             & \E \sup_{u\in \Hscr(1)} \left\|   (\mathcal{S} + \lambda \mathcal{I})^{-1/2} \left[\frac{1}{n}\sum_{i=1}^n \mathcal{K}_{A_i} \left\{\truew_i  u(A_i, \bm X_i) - z(A_i;u)\right\}\right]\right\|^2_{\Hscr_A} \nonumber\\
              \leq & 4\E  \sup_{u\in \Hscr(1)}  \left\|   (\mathcal{S} + \lambda \mathcal{I})^{1/2} \frac{1}{n}\sum_{i=1}^n K_{A_i} \{r_i\truew_i u(A_i, \bm  X_i)\}\right\|^2_{\Hscr_A}\nonumber\\
              & \qquad \qquad \qquad + 4\E  \sup_{u\in \Hscr(1)}  \left\|   (\mathcal{S} + \lambda  \mathcal{I})^{1/2} \frac{1}{n}\sum_{i=1}^n K_{A_i}r_i z(A_i;u)\right\|^2_{\Hscr_A} \label{eqn:radecomp}
            \end{align}
            
            Let's focus on the first term in \eqref{eqn:radecomp}.
            \begin{align}
              &\E  \sup_{u\in \Hscr(1)}  \left\|   (\mathcal{S} + \lambda \mathcal{I})^{1/2} \frac{1}{n}\sum_{i=1}^n \mathcal{K}_{A_i} \{r_i\truew_i u(A_i,\bm   X_i)\}\right\|^2_{\Hscr_A} \nonumber\\
              = &
              \E \sup_{u\in \Hscr(1)} \left\langle (\mathcal{S} + \lambda \mathcal{I})^{-1/2}\frac{1}{n}\sum_{i=1}^n \mathcal{K}_{A_i} \{r_i \truew_i u(A_i, \bm  X_i)\},\right.\nonumber\\
              &\qquad \qquad \qquad \qquad \left. (\mathcal{S} + \lambda \mathcal{I})^{-1/2} \frac{1}{n}\sum_{i=1}^n \mathcal{K}_{A_i} \{r_i \truew_i u(A_i, \bm 
   X_i) \}\right\rangle_{\Hscr_A} \nonumber\\
              =   & \E \sup_{u\in \Hscr(1)} \frac{1}{n^2} \sum_{i=1}^n \sum_{j=1}^n r_i r_j  \truew_i \truew_j u(A_i, \bm  X_i) u(A_j, \bm  X_j)\nonumber\\
               &\qquad \qquad\qquad \times \left\langle (\mathcal{S} + \lambda \mathcal{I})^{-1/2} \mathcal{K}_{A_i}, (\mathcal{S} + \lambda \mathcal{I})^{-1/2}\mathcal{K}_{A_j} \right\rangle_{\Hscr_A}\nonumber\\
              \leq &  \frac{1 }{n^2} \E \sup_{u\in \Hscr(1)} \sum_{i=1}^n (\truew_i)^2 u^2(A_i, \bm 
   X_i) \left\langle (\mathcal{S} + \lambda \mathcal{I})^{-1/2} \mathcal{K}_{A_i}, (\mathcal{S} + \lambda \mathcal{I})^{-1/2}\mathcal{K}_{A_i} \right\rangle_{\Hscr_A} \nonumber\\ 
              +& \frac{1}{n^2}\E \sup_{u\in \Hscr(1)} \sum_{i\neq j}  r_i r_j \truew_i \truew_j u(A_i, \bm  X_i) u(A_j, \bm  X_j) \left\langle (\mathcal{S} + \lambda \mathcal{I})^{-1/2} \mathcal{K}_{A_i}, (\mathcal{S} + \lambda \mathcal{I})^{-1/2}\mathcal{K}_{A_j} \right\rangle_{\Hscr_A} \nonumber \\
              & = (i) + (ii) \label{eqn:redacomp1_decomp}
             \end{align}

            We first deal with the (i).
            For every \(u\in \mathcal{H}(1)\),  \(\|u\|_\infty  = \sup_{(a,\bm x)} |\langle\kernel((a,\bm x), (\cdot,\cdot)), u\rangle_{\Hscr}|\leq  \sup_{(a,x)} |\kernel((a,\bm x), (a,\bm x))| \leq \oldu{cnt:kernelbound}\kappa\). By contraction inequality and symmetrization inequality, we have
            \begin{align*}
              &\E \sup_{u\in \Hscr(1)} \sum_{i=1}^n (\truew_i)^2 u^2(A_i, \bm X_i) \left\langle (\mathcal{S} + \lambda \mathcal{I})^{-1/2} \mathcal{K}_{A_i}, (\mathcal{S} + \lambda \mathcal{I})^{-1/2}\mathcal{K}_{A_i} \right\rangle_{\Hscr_A} \\
              \leq &n \oldu{cnt:kernelbound}^2\kappa^2 \E \left\langle (\mathcal{S} + \lambda \mathcal{I})^{-1/2} K_{A_1}, (\mathcal{S} + \lambda \mathcal{I})^{-1/2}\mathcal{K}_{A_1} \right\rangle_{\Hscr_A}\nonumber\\
              \leq & n \oldu{cnt:kernelbound}^2\kappa^2 \E \|\mathcal{K}_{A_1}^*(\mathcal{S} + \lambda \mathcal{I})^{-1}\mathcal{K}_{A_1}\|_{\mathcal{L}(\Hscr_A)}\nonumber\\
              \leq & n \oldu{cnt:kernelbound}^2\kappa^2 \E \{\mbox{Tr}(\mathcal{K}_{A_1}^*(\mathcal{S}+ \lambda \mathcal{I})^{-1}\mathcal{K}_{A_1})\}\\
               =&  n \oldu{cnt:kernelbound}^2\kappa^2 \E \{\mbox{Tr}((\mathcal{S} + \lambda \mathcal{I})^{-1}\mathcal{K}_{A_1}\mathcal{K}_{A_1}^*)\}
              =  n \oldu{cnt:kernelbound}^2\kappa^2  \int_a \mbox{Tr}\left\{(\mathcal{S} + \lambda \mathcal{I})^{-1}\mathcal{S}_a \right\}d \rho_A(a) \nonumber\\
              \leq & n \oldu{cnt:kernelbound}^2\kappa^2 \mbox{Tr}\left\{(\mathcal{S} + \lambda \mathcal{I})^{-1}\mathcal{S} \right\} = n \oldu{cnt:kernelbound}^2\kappa^2 \mathcal{N}(\lambda).
            \end{align*}
            Next, we study the term  (ii).
            Based on the proof in Proposition \ref{prop:tau}, we have \(\kernel_A(\cdot, \cdot) = \sum_{l_1=1}^\infty \nu_{l_1}^A \phi_{l_1}^A(\cdot) \phi_{l_1}^A(\cdot) \), where \(\nu_{l_1}^A\) are eigenvalues and \(\phi_{l_1}^A(\cdot)\) are eigenfunctions (orthonormal basis of \(\mathcal{L}_2(\rho_A)\)). \(\kernel_X(\cdot, \cdot) = \sum_{l_2=1}^\infty \nu_{l_2}^X \phi_{l_2}^X(\cdot) \phi_{l_2}^X(\cdot) \), where \(\nu_{l_2}^X\) are eigenvalues and \(\phi_{l_2}^X(\cdot)\) are eigenfunctions (orthonormal basis of \(\mathcal{L}_2(\rho_X)\). 
            Since the reproducing kernel for \(\Hscr\) is \(\kernel((\cdot, \star),(\cdot, \star)) = \kernel_A(\cdot, \cdot)\kernel_X(\star, \star) \), we have \(\kernel((\cdot, \star),(\cdot, \star)) = \sum_{l=1}^\infty \nu_l \phi_{l}(\cdot,\star)  \phi_{l}(\cdot,\star)\), where \(\nu_l = \nu_{l_1}^A \nu_{l_2}^X\) and \(\phi_{l}(\cdot,\star)  = \phi_{l_1}^A(\cdot)\phi_{l_2}^X(\star)\) for some \(l_1, l_2\) such that \(\nu\), \(l= 1,\dots, \infty\) is nonincreasing.
            Take \(\Phi(\cdot, \star) = \{\sqrt{ \nu} \phi_l(\cdot, \star)\}_{l=1}^\infty\). \(\Hscr(1) = \{u(\cdot, \star) = \langle \beta, \Phi(\cdot, \star)\rangle : \sum_{l=1}^\infty \beta^2_l \leq 1\}\).  
            Take \(\mathcal{E}(1) = \{\beta:\sum_{l=1}^\infty \beta^2_l  \leq 1 \}\), then
            \begin{align*}
              &(ii) \leq \E  \sup_{\beta \in \mathcal{E}(1)} \left\{ \sum_{l=1}^\infty (\beta_l )^2\sum_{l'=1}^\infty (\beta_{l'} )^2\right\}^{\frac{1}{2}}\\
              & \Scale[0.85]{\times\left[\sum_{l=1}^\infty \sum_{l'=1}^\infty  \left\{  \sum_{i \neq j} \left\langle (\mathcal{S} + \lambda \mathcal{I})^{-\frac{1}{2}} \mathcal{K}_{A_i}, (\mathcal{S} + \lambda \mathcal{I})^{-\frac{1}{2}}\mathcal{K}_{A_j} \right\rangle_{\Hscr_A}  r_i r_j\truew_i\truew_j \sqrt{\nu_l}\phi_l(A_i, \bm X_i) \sqrt{\nu_{l'}}\phi_{l'}(A_j, \bm  X_j)\right\}^2 \right]^{\frac{1}{2}}}\\
              \leq & \Scale[0.9]{\left[ \sum_{l=1}^\infty \sum_{l'=1}^\infty   \sum_{i \neq j} \E   \left\langle (\mathcal{S} + \lambda \mathcal{I})^{-\frac{1}{2}} \mathcal{K}_{A_i}, (\mathcal{S} + \lambda \mathcal{I})^{-\frac{1}{2}}\mathcal{K}_{A_j} \right\rangle_{\Hscr_A}^2 (\truew_i)^2 (\truew_j)^2 \nu_l\phi^2_l(A_i, \bm X_i)\nu_{l'}\phi^2_{l'}(A_j, \bm X_j) \right]^{\frac{1}{2}}}\\
              \leq & \left[\sum_{l=1}^\infty \sum_{l'=1}^\infty \sum_{i\neq j}\E \left\{   (w_i^*)^2  \left\langle (\mathcal{S} + \lambda \mathcal{I})^{-\frac{1}{2}} K_{A_i}, (\mathcal{S} + \lambda \mathcal{I})^{-\frac{1}{2}}K_{A_i} \right\rangle_{\Hscr_A} \nu_l\phi_l^2(A_i,\bm  X_i)\right\} \right.\\
              &\left. \E \left\{  (w_j^*)^2 \left\langle (\mathcal{S} + \lambda \mathcal{I})^{-\frac{1}{2}} K_{A_j}, (\mathcal{S} + \lambda \mathcal{I})^{-\frac{1}{2}}K_{A_j} \right\rangle_{\Hscr_A} \nu_{l'}\phi_{l'}^2(A_j,\bm  X_j)\right\}\right]^{\frac{1}{2}}\\
              = & \left[ \sum_{i\neq j}\E \left\{   (w_i^*)^2  \left\langle (\mathcal{S} + \lambda \mathcal{I})^{-\frac{1}{2}} K_{A_i}, (\mathcal{S} + \lambda \mathcal{I})^{-\frac{1}{2}}K_{A_i} \right\rangle_{\Hscr_A} \left( \sum_{l=1}^\infty \nu_l\phi_l^2(A_i, \bm X_i)\right)\right\}  \right.\\
              &\left.\E \left\{  (w_j^*)^2 \left\langle (\mathcal{S} + \lambda \mathcal{I})^{-\frac{1}{2}} K_{A_j}, (\mathcal{S} + \lambda \mathcal{I})^{-\frac{1}{2}}K_{A_j} \right\rangle_{\Hscr_A} \left(\sum_{l'=1}^\infty \nu_{l'}\phi_{l'}^2(A_j,\bm  X_j) \right)\right\}\right]^{\frac{1}{2}}.
            \end{align*}
            The first inequality by adopting Cauchy Schwarz inequality. The second inequality is due to that $r_i, \  i = 1,\dots, n$  are all independent Rademacher random variables. The third inequality is because that 
            $(A_i, X_i),\  i = 1,\dots, n$ are independent pairs and 
             \begin{align*}
                &\left\langle (\mathcal{S} + \lambda \mathcal{I})^{-1/2} \mathcal{K}_{A_i}, (\mathcal{S} + \lambda \mathcal{I})^{-1/2}\mathcal{K}_{A_j} \right\rangle_{\Hscr_A}^2 \\
                \leq & \left\langle (\mathcal{S} + \lambda \mathcal{I})^{-1/2} \mathcal{K}_{A_i}, (\mathcal{S} + \lambda \mathcal{I})^{-1/2}\mathcal{K}_{A_i} \right\rangle_{\Hscr_A} \left\langle (\mathcal{S} + \lambda \mathcal{I})^{-1/2} \mathcal{K}_{A_j}, (\mathcal{S} + \lambda \mathcal{I})^{-1/2}\mathcal{K}_{A_j} \right\rangle_{\Hscr_A}
             \end{align*} 
            Note that 
            \begin{align}
             \sum_{l=1}^\infty \nu_l\phi_l^2(A_i, \bm X_i)  = \kernel_A(A_i,A_i)\kernel_X(\bm X_i, \bm X_i) \leq \kappa \oldu{cnt:kernelbound}. \nonumber
            \end{align}
            Then 
            \begin{align}
              &\E \left\{   (w_i^*)^2  \left\langle (\mathcal{S} + \lambda \mathcal{I})^{-1/2} K_{A_i}, (\mathcal{S} + \lambda \mathcal{I})^{-1/2}K_{A_i} \right\rangle_{\Hscr_A} \left( \sum_{l=1}^\infty \nu_l\phi_l^2(A_i, X_i)\right)\right\}  \nonumber\\
              \leq& \oldu{cnt:overlapupper}^2 \kappa \oldu{cnt:kernelbound} \E \left\langle (\mathcal{S} + \lambda \mathcal{I})^{-1/2} K_{A_i}, (\mathcal{S} + \lambda \mathcal{I})^{-1/2}K_{A_i} \right\rangle_{\Hscr_A} \nonumber \\
              \leq &   \oldu{cnt:overlapupper}^2\kappa \oldu{cnt:kernelbound} \mathcal{N}(\lambda)\nonumber
            \end{align}
            
            Now we prove that 
            \begin{align}
              (ii) \leq \frac{\oldu{cnt:overlapupper}^2 \kappa \oldu{cnt:kernelbound} }{n^2}\left\{ \sqrt{n(n-1)} \mathcal{N}(\lambda)\right\} \leq \frac{\oldu{cnt:overlapupper}^2 \kappa \oldu{cnt:kernelbound}}{n} \mathcal{N}(\lambda). \nonumber
            \end{align}
            Combine the bound of \((i)\) and \((ii)\) into \eqref{eqn:redacomp1_decomp}, we have
            \begin{align}
                \E  \sup_{u\in \Hscr(1)}  \left\|   (\mathcal{S} + \lambda \mathcal{I})^{1/2} \frac{1}{n}\sum_{i=1}^n \mathcal{K}_{A_i} \{r_i\truew_i u(A_i, \bm X_i)\}\right\|^2_{\Hscr_A} \leq \frac{ \oldu{cnt:kernelbound}^2\kappa^2 \mathcal{N}(\lambda)}{n} +  \frac{\oldu{cnt:overlapupper}^2 \kappa \oldu{cnt:kernelbound}}{n} \mathcal{N}(\lambda) 
            \end{align}
            
            Next we deal with the second term in \eqref{eqn:radecomp}.
            For \(u \in \Hscr(1)\), by previous construction of \(\Hscr(1)\), we can express \(u(a,x) = \sum_{l_1}\sum_{l_2}\beta^A_{l_1}\beta^X_{l_2}\sqrt{\nu^A_{l_1} \nu^X_{l_2}} \phi_{l_1}^A(a) \phi_{l_2}^X (x)\) for some coefficients \(\beta^A_{l_1}\) and \(\beta^X_{l_2}\). Then 
            \begin{align*}
                \|z(\cdot;u)\|^2_{\Hscr_A} &= \sum_{l_1} (\beta^A_{l_1})^2 \left\{\E \left(  \sum_{l_2} \sqrt{\nu_{l_2}^X}\beta^X_{l_2}\phi_{l_2}^X(\bm X)  \right)\right\}^2
                \leq \sum_{l_1} (\beta^A_{l_1})^2\E \left(  \sum_{l_2} \sqrt{\nu_{l_2}^X}\beta^X_{l_2}\phi_{l_2}^X(\bm X)  \right)^2\\
            & \leq \sum_{l_1} (\beta^A_{l_1})^2 \sum_{l_2}\nu_{l_2}^X (\beta_{l_2}^X)^2
             \leq \left( \max_{l_2} \nu_{l_2}^X \right) \sum_{l_1}\sum_{l_2} (\beta^A_{l_1})^2(\beta_{l_2}^X)^2 = \nu_{1}^X
            \end{align*}
            
            Then \(\{z(\cdot;u) : \|u\|_\Hscr \leq 1\} \subset\{z: \|z\|_{\Hscr_A} \leq \nu_1^X\}\), we can follow previous strategy and prove that 
            \begin{align}
                &\E  \sup_{u\in \Hscr(1)}  \left\|   (\mathcal{S} + \lambda \mathcal{I})^{1/2} \frac{1}{n}\sum_{i=1}^n K_{A_i}r_i z(A_i;u)\right\|^2_{\Hscr_A}\\
                &\leq  (\nu_1^X)^2 \E \sup_{z\in \{z: \|z\|_{\Hscr_A} \leq 1\}} \left\|   (\mathcal{S} + \lambda \mathcal{I})^{1/2} \frac{1}{n}\sum_{i=1}^n K_{A_i}r_i z(A_i)\right\|^2_{\Hscr_A}\nonumber\\
                & \leq \frac{  (\nu_{1}^X)^2 \kappa^2 \mathcal{N}(\lambda)}{n} +  \frac{( \nu_{1}^X)^2 \oldu{cnt:overlapupper}^2 \kappa }{n} \mathcal{N}(\lambda) \nonumber
            \end{align}

            And overall, there exists a constant \(\newl\ltxlabel{cnt:exp_L2_bound}>0\) depending on \(\nu_1^X, \kappa,\oldu{cnt:overlapupper}, \oldu{cnt:kernelbound} \), such that 
            \begin{align}
                \E \sup_{u\in \Hscr(1)} \left\|   (\mathcal{S} + \lambda \mathcal{I})^{-1/2} \left[\frac{1}{n}\sum_{i=1}^n \mathcal{K}_{A_i} \left\{\truew_i  u(A_i, \bm X_i) - z(A_i;u)\right\}\right]\right\|^2_{\Hscr_A} \leq \frac{\oldl{cnt:exp_L2_bound}\mathcal{N}(\lambda)}{n}. \nonumber
            \end{align}
            And  combine the result with \eqref{eqn:opnorm_bound}, we have \[\sup_{u\in \Hscr(1)} \left\| \left( \frac{1}{n}\sum_{i=1}^n \mathcal{S}_{A_i} + \lambda  \mathcal{I}\right)^{-1}\left[\frac{1}{n}\sum_{i=1}^n \mathcal{K}_{A_i} \left\{\truew_i  u(A_i, \bm X_i) - z(A_i;u)\right\}\right] \right\|^2_{n} = \bigOp\left( \frac{\mathcal{N}(\lambda)}{n}\right).\]

            \item Next, we consider to bound \eqref{eqn:comp2}. By the above arguments,
         \begin{align}
         & \sup_{u\in \Hscr(1)} \left\| \left( \frac{1}{n}\sum_{i=1}^n \mathcal{S}_{A_i} + \lambda  \mathcal{I}\right)^{-1}\left[\frac{1}{n}\sum_{i=1}^n \mathcal{K}_{A_i}  z(A_i;u)\right] - z(\cdot;u) \right\|_{n}^2 \nonumber \\
           \leq& \nu_1^X \sup_{z\in \{z:\|z\|_{\Hscr_A} \leq 1\}} \left\| \left( \frac{1}{n}\sum_{i=1}^n \mathcal{S}_{A_i} + \lambda  \mathcal{I}\right)^{-1}\left[\frac{1}{n}\sum_{i=1}^n \mathcal{K}_{A_i}  z(A_i)\right] - z \right\|_{n}^2 \nonumber
         \end{align}
         Take \(z^{\lambda} = (\mathcal{S} + \lambda \mathcal{I})^{-1} S z\).  We can verify that  
  \begin{align*}
     \| z^{\lambda} - z\|^2_{\mathcal{H}_A} &= \left\| \left[ (\mathcal{S} + \lambda \mathcal{I})^{-1} S - \mathcal{I} \right] z\right\|^2_{\mathcal{H}_A} 
     = \sum_{l=1}^\infty \left[ \frac{\nu_l^A}{\nu_l^A + \lambda} - 1 \right]^2 \langle z, \phi_l^A\rangle_{\mathcal{H}_A}^2 \leq \sum_{l=1}^\infty   \langle z, \phi_l^A\rangle_{\mathcal{H}_A}^2  = \|z\|_{\mathcal{H}_A}^2.\\
      \| z^{\lambda} - z\|^2_{\mathcal{L}_2({\rho_A})} &= \left\| \sqrt{\mathcal{S}} \left[ (\mathcal{S} + \lambda \mathcal{I})^{-1} S \right]z  -  z\right\|^2_{\mathcal{H}_A}  = \sum_{l=1}^\infty \nu_l^A\left[ \frac{\nu_l^A}{\nu_l^A + \lambda} - 1 \right]^2 \langle z, \phi_l^A\rangle_{\mathcal{H}_A}^2 \\
     & = \sum_{l=1}^\infty \left[ \frac{\lambda}{\sqrt{\nu_l^A} + \lambda/\sqrt{\nu_l^A}} \right]^2  \langle z, \phi_l^A\rangle_{\mathcal{H}_A}^2 
      \leq \sum_{l=1}^\infty  \left(\frac{\lambda}{\sqrt{2\lambda} }  \right)^2\langle z, \phi_l^A\rangle_{\mathcal{H}_A}^2 = \lambda/2 \|z\|_{\mathcal{H}_A}^2.
  \end{align*}
         Next, we derive the bound for
         \begin{align*}
            \sup_{z \in \{z: \|z\|_{\Hscr_A} \leq 1\}} \|z^{\lambda} - z\|_n^2.
          \end{align*}
          Take \(z' = z^\lambda - z\),  given  that 
          \(\|z'\|_{\Hscr_A} \leq \sqrt{\newl\ltxlabel{cnt:zhnorm}} \|z\|_{\Hscr_A} \leq \sqrt{\oldl{cnt:zhnorm}}\) and \(\|z'\|_{\mathcal{L}_2(\rho_A)} \leq \sqrt{\oldl{cnt:zl2norm}\lambda} \|z\|_{\Hscr_A} \leq \sqrt{ \newl\ltxlabel{cnt:zl2norm}\lambda}\) for some positive constants \oldl{cnt:zhnorm} and \oldl{cnt:zl2norm}. 
          Follow the same proof of Lemma 42 in \cite{mendelson2002geometric}, we can show that the Rademacher complexity 
          \begin{align*}
            \E \sup_{z'\in \{z: \|z\|_{\Hscr_A} \leq \sqrt{\oldl{cnt:zhnorm}}, \|z\|_{\mathcal{L}_2(\rho_A)} \leq \sqrt{\oldl{cnt:zl2norm}\lambda}\}}\left| \frac{1}{n}\sum_{i=1}^n r_i z'(A_i) \right|^2 \leq \frac{\newl\ltxlabel{cnt:emp_rade}}{n} \left( \sum_{l=1}^\infty \min \{\nu^A_l,\lambda\} \right)
          \end{align*}
          for some constant \(\oldl{cnt:emp_rade}>0\) depending on \(\oldl{cnt:zhnorm}\) and \(\oldl{cnt:zl2norm}\).
          Next, we apply Corollary 2.2 in \cite{bartlett2005local}, we can verify there exists a constant \(b>0\) such that  \(\|z'\|_\infty \leq b\) for any \(\|z'\|_{\Hscr_A} \leq \sqrt{\oldl{cnt:zhnorm}}\). Then for any \(x > 0\),  if \(\lambda \ge 10 b \{{\oldl{cnt:emp_rade}}  \sum_{l=1}^\infty \min \{\nu^A_l,\lambda\} /n\}^{1/2} + 11b^2x/n\), we have 
          \begin{align*}
            \{z'\in \{z: \|z\|_{\Hscr_A} \leq \sqrt{\oldl{cnt:zhnorm}}\} :  \|z'\|^2_{\mathcal{L}_2(\rho_A)} \leq  \oldl{cnt:zl2norm}\lambda \} \subseteq \{z'\in \{z: \|z\|_{\Hscr_A} \leq \sqrt{\oldl{cnt:zhnorm}}\} :  \|z'\|^2_{n} \leq 2 \oldl{cnt:zl2norm}\lambda \},
          \end{align*}
          with probability at least \(1-\exp(-x)\). Note that \(v^A_l = t_l\), then as long as \(\sqrt{ \sum_{l=1}^\infty \min \{t_l,\lambda\}}/(\sqrt{n}\lambda) = \bigOp(1)\), we have \(\|z'\|_n^2 = \bigOp(\lambda)\). The above inequalities also holds for \(\hat A_i\), \(i = 1,\dots, n\) as $\hat A_i$ are independent samples from $\mathcal{A}$.  
         Then we have the following the inequality
         \begin{align}
         & \sup_{z\in \{z:\|z\|_{\Hscr_A} \leq 1\}} \left\| \left( \frac{1}{n}\sum_{i=1}^n \mathcal{S}_{A_i} + \lambda  \mathcal{I}\right)^{-1}\left[\frac{1}{n}\sum_{i=1}^n \mathcal{K}_{A_i}  z(A_i)\right] - z \right\|_{n}^2 \nonumber\\
          \leq &2\sup_{z\in \{z:\|z\|_{\Hscr_A} \leq 1\}} \left\| \left( \frac{1}{n}\sum_{i=1}^n \mathcal{S}_{A_i} + \lambda  \mathcal{I}\right)^{-1} \left( \mathcal{S} - \frac{1}{n}\sum_{i=1}^n \mathcal{S}_{A_i} \right) \left( z^{\lambda} - z \right) \right\|_{n}^2 \nonumber\\
          &+  2\sup_{z\in \{z:\|z\|_{\Hscr_A} \leq 1\}} \| z^{\lambda} - z\|^2_{n}\nonumber\\
          \leq & \left\|\left( \frac{1}{n}\sum_{i=1}^n \mathcal{S}_{\hat A_i} \right)^{1/2} \left( \frac{1}{n}\sum_{i=1}^n \mathcal{S}_{A_i} + \lambda  \mathcal{I}\right)^{-1} (S + \lambda \mathcal{I})^{\frac{1}{2}} \right\|^2_{\mathcal{L} (\Hscr_A)} \nonumber\\
          & \qquad  \oldl{cnt:zhnorm}\sup_{v \in \{z:\|z\|_{\Hscr_A} \leq 1\}}\left\| \left(\mathcal{S} + \lambda  \mathcal{I}\right)^{-\frac{1}{2}} \left( \mathcal{S} - \frac{1}{n}\sum_{i=1}^n \mathcal{S}_{A_i} \right) v \right\|_{\Hscr_A}^2  + \oldl{cnt:zl2norm}\lambda \label{eqn:comp2_comp}
         \end{align}
       The operator norm in \eqref{eqn:comp2_comp} can be bounded via \eqref{eqn:opnorm_bound}. It remains to bound \[\sup_{v \in \{z:\|z\|_{\Hscr_A} \leq 1\}}\left\| \left(\mathcal{S} + \lambda  \mathcal{I}\right)^{-\frac{1}{2}} \left( \mathcal{S} - \frac{1}{n}\sum_{i=1}^n \mathcal{S}_{A_i} \right) v \right\|^2_{\Hscr_A}.\]

       Note that \(\E \mathcal{S}_{A_i} = \mathcal{S}\), then we can apply the symmetrization equality again 
       \begin{align*}
         \sup_{v \in \{z:\|z\|_{\Hscr_A} \leq 1\}}\left\| \left(\mathcal{S} + \lambda  \mathcal{I}\right)^{-\frac{1}{2}} \left( \mathcal{S} - \frac{1}{n}\sum_{i=1}^n \mathcal{S}_{A_i} \right) v \right\|^2_{\Hscr_A} 
         \leq 4 \E \sup_{v \in \{z:\|z\|_{\Hscr_A} \leq 1\}}\left\| \left(\mathcal{S} + \lambda  \mathcal{I}\right)^{-\frac{1}{2}} \frac{1}{n}\sum_{i=1}^n \mathcal{S}_{A_i}r_i v \right\|^2_{\Hscr_A},
       \end{align*}
       where \(r_i\), \(i = 1,\dots, n\) are independent Rademacher random variables.
    
       Follow the similar proof in proving term \eqref{eqn:comp1}, we can show that 
       \begin{align*}
         \E \sup_{v \in \{z:\|z\|_{\Hscr_A} \leq 1\}}\left\| \left(\mathcal{S} + \lambda  \mathcal{I}\right)^{-\frac{1}{2}} \frac{1}{n}\sum_{i=1}^n \mathcal{S}_{A_i}r_i v \right\|^2_{\Hscr_A} \leq \newl\ltxlabel{cnt:rade2} \frac{\mathcal{N}(\lambda)}{n}
       \end{align*}
       for some constant \(\oldl{cnt:rade2}>0\) depending on \(\kappa\). And overall, we prove that 
       \begin{align*}
         \eqref{eqn:comp2} = \bigOp\left( \frac{\mathcal{N}(\lambda)}{n} + \lambda \right)
       \end{align*}
    
    \item  Last, we bound the term \eqref{eqn:comp3}.
    First, fixed any \(a \in \mathcal{A}\), we derive the bound for \(\E  \sup_{u\in \Hscr(1)}  [  z(a;u) -  \sum_{j=1}^n u(a, \bm X_j)/n]^2 \).
    
    By symmetrization inequality, we have 
    \begin{align*}
      \E  \sup_{u\in \Hscr(1)}  \left[  z(a;u) -  \frac{\sum_{j=1}^n u(a, \bm X_j)}{n} \right] ^2 \leq 4 \E \sup_{u\in \Hscr(1)}\left[ \frac{1}{n}\sum_{j=1}^n r_j u(a, \bm X_j) \right] ^2 
    \end{align*}
    Again by previous construction of \(\Hscr(1)\), we have 
    \(\Hscr(1) = \{u(\cdot, \star) = \langle \beta, \Phi(\cdot, \star)\rangle : \sum_{l=1}^\infty \beta^2_l \leq 1\}\). Take  \(\mathcal{E}(1) = \{\beta:\sum_{l=1}^\infty \beta^2_l  \leq 1 \}\). 
    \begin{align*}
      &\E \sup_{u\in \Hscr(1)}\left[ \frac{1}{n}\sum_{j=1}^n r_i u(a, \bm X_j) \right] ^2  
    \leq  \E \sup_{\beta \in\mathcal{E}(1)} \left[ \frac{1}{n}\sum_{j=1}^n r_j \left\{ \sum_{l}\beta_l\sqrt{\nu_l}\phi_l(a,\bm  X_j) \right\} \right] ^2 \\
    \leq & \E \sup_{\beta \in\mathcal{E}(1)}  \left( \sum_{l}\beta_l^2 \right) \left[ \frac{1}{n}\sum_{j=1}^n r_j \sqrt{\nu_l} \phi_l(a, \bm X_j) \right]^2 
    \leq  \sum_l \nu_l \E \left[ \frac{1}{n}\sum_{j=1}^n r_j \phi_l(a, \bm X_j) \right]^2 \leq \frac{1}{n}\sum_l \nu_l \E \phi_l^2(a,\bm X_1)\\
    = & \frac{1}{n}\kernel_A(a,a)\E \kernel_X(\bm X_1, \bm X_1) \leq \frac{1}{n}\kappa \oldu{cnt:kernelbound}\kernel_A(a,a).
    \end{align*}
    Then we can show that
    \begin{align*}
        &\E \left\{ \sup_{u\in \Hscr(1)} \left\| z(\cdot;u) -  \frac{1}{n}\sum_{j=1}^n u(\cdot, \bm X_j)\right\|_{n}^2 \right\} \leq \E \left( \E \left\{  \sup_{u\in \Hscr(1)}  \left[z(A;u) -  \frac{1}{n}\sum_{j=1}^n u(A, \bm X_j)\right]^2 \mid A \right\}  \right)\\
        & \lesssim  \frac{1}{n}\kappa \kernel_A(A,A) \lesssim \frac{1}{n}= \bigOp\left( \frac{1}{\sqrt{n}} \right).
    \end{align*}
    
    Combine the bounds derived for \eqref{eqn:comp1}, \eqref{eqn:comp2} and \eqref{eqn:comp3}, the bound for \(\sup_{u \in\Hscr(1)}Q(\truew, \lambda, \|\cdot\|_n, u)\) follows. 
    
    To bound the penalty term \(R(\truew, \lambda)\), note that 
    \begin{align*}
     & R(\bm \truew, \lambda,) = \frac{1}{n^2} \sum_{i=1}^n (\truew_i)^2 \left\| \left( \frac{1}{n}\sum_{i=1}^n \mathcal{S}_{\hat A_i} + \lambda  \mathcal{I}\right)^{-1}\mathcal{K}_{\hat A_i} \right\|_{n}^2\\
      &\leq \frac{\oldu{cnt:overlapupper}}{n^2} \sum_{i=1}^n  \left\| \left( \frac{1}{n}\sum_{i=1}^n \mathcal{S}_{\hat A_i} + \lambda \mathcal{I} \right)^{-1}\mathcal{K}_{\hat A_i} \right\|_{n}^2\\
    &\leq  \frac{\oldu{cnt:overlapupper}}{n^2} \sum_{i=1}^n  \left\|\left( \frac{1}{n}\sum_{i=1}^n \mathcal{S}_{\hat A_i} \right)^{1/2} \left( \frac{1}{n}\sum_{i=1}^n \mathcal{S}_{\hat A_i} + \lambda \mathcal{I} \right)^{-1} (\mathcal{S} + \lambda \mathcal{I})^{\frac{1}{2}} \right\|^2_{\mathcal{L} (\Hscr_A)}  \left\| (\mathcal{S} + \lambda \mathcal{I})^{-\frac{1}{2}} \mathcal{K}_{\hat A_i}\right\|^2_{\Hscr_A}.
    \end{align*}
    \begin{align*}
      &  \left\|\left( \frac{1}{n}\sum_{i=1}^n \mathcal{S}_{\hat A_i} \right)^{1/2} \left( \frac{1}{n}\sum_{i=1}^n \mathcal{S}_{\hat A_i} + \lambda  \mathcal{I}\right)^{-1} (\mathcal{S} + \lambda \mathcal{I})^{\frac{1}{2}} \right\|_{\mathcal{L} (\Hscr_A)}  \\
         \leq &  \Bigg\|\left( \frac{1}{n}\sum_{i=1}^n \mathcal{S}_{\hat A_i} \right)^{1/2}\left( \frac{1}{n}\sum_{i=1}^n \mathcal{S}_{\hat A_i} + \lambda  \mathcal{I}\right)^{-1} \left( \frac{1}{n}\sum_{i=1}^n \mathcal{S}_{\hat  A_i} -  \frac{1}{n}\sum_{i=1}^n \mathcal{S}_{ A_i}\right)\\
         &\qquad\qquad\qquad \times (\mathcal{S}_{\bm A} + \lambda \mathcal{I})^{-1}  (\mathcal{S} + \lambda \mathcal{I})^{\frac{1}{2}} \Bigg\|_{\mathcal{L} (\Hscr_A)}  \\
         & + \left\|\left( \frac{1}{n}\sum_{i=1}^n \mathcal{S}_{\hat A_i} \right)^{1/2} \left( \frac{1}{n}\sum_{i=1}^n \mathcal{S}_{ A_i} + \lambda \mathcal{I} \right)^{-1} (\mathcal{S} + \lambda \mathcal{I})^{\frac{1}{2}} \right\|_{\mathcal{L} (\Hscr_A)}  \\
         & = \bigOp(\lambda^{-1} \kappa^h ) + \bigOp(1) = \bigOp(1).
    \end{align*}
    The last two equalities are due to \eqref{eqn:opnorm_bound} and the condition of $\kappa$.
  \begin{align*}
    \left\| (\mathcal{S} + \lambda \mathcal{I})^{-\frac{1}{2}} \mathcal{K}_{\hat A_i}\right\|^2_{\Hscr_A}  = \left\|  (\mathcal{S} + \lambda \mathcal{I})^{-\frac{1}{2}} \left( \mathcal{S}_{\hat A_i} - \mathcal{S}_{A_i}\right)(\mathcal{S} + \lambda \mathcal{I})^{-\frac{1}{2}}\right\|_{\mathcal{L}(\Hscr_A)} \\
   + \left\|  (\mathcal{S} + \lambda \mathcal{I})^{-\frac{1}{2}} \mathcal{S}_{A_i}(\mathcal{S} + \lambda \mathcal{I})^{-\frac{1}{2}}\right\|_{\mathcal{L}(\Hscr_A)}\\
     \leq \left\|  \left( \mathcal{S}_{\hat A_i} - \mathcal{S}_{A_i}\right)(\mathcal{S} + \lambda \mathcal{I})^{-1}\right\|_{\mathcal{L}(\Hscr_A)} + \left\|  \mathcal{S}_{A_i} (\mathcal{S} + \lambda \mathcal{I})^{-1}\right\|_{\mathcal{L}(\Hscr_A)} \\
     \leq  \frac{1}{\lambda}\left\| \mathcal{S}_{\hat A_i} - \mathcal{S}_{A_i}\right\|_{\mathcal{L}(\Hscr_A)} +  \left\|  \mathcal{S}_{A_i} (\mathcal{S} + \lambda \mathcal{I})^{-1}\right\|_{\mathcal{L}(\Hscr_A)} 
  \end{align*}
  
    It's easy to verify that 
    \begin{align*}
     \E \| \mathcal{S}_{A_i} (\mathcal{S} + \lambda \mathcal{I})^{-1} \|_{\mathcal{L}(\Hscr_A)} 
     \leq  \E \left\{ \mbox{Tr} \left( (\mathcal{S} + \lambda \mathcal{I})^{-1} \mathcal{K}_{ A_i}  \mathcal{K}_{ A_i}^*\right) \right\} \leq  \mbox{Tr}\{(\mathcal{S}+ \lambda \mathcal{I})^{-1} \mathcal{S}\}  = \mathcal{N}(\lambda).
    \end{align*}
    Then combine with all the bounds, we obtain \[R(\bm \truew, \lambda) = \bigOp\left( \frac{\mathcal{N}(\lambda)}{n} \right) + \bigOp\left( \frac{\kappa^h}{\lambda}  \right) \leq \bigOp\left( \frac{\mathcal{N}(\lambda)}{n} \right).\]
        \end{itemize}
        The last inequality is due to the condition of $\kappa$.
  
  Now we are ready to bound $\sup_{u \in \Hscr(1)} Q(\hat{\bm w}, \lambda, u)$ and $R(\hat{\bm w}, \lambda)$.
   Since \(\bm \estw\) is the solution of \eqref{eqn:estw_def}, by the basic inequality, we have 
      \begin{align*}
        \sup_{u \in \Hscr(1)} Q(\bm \estw, \lambda,  u) + \eta R(\bm \estw, \lambda) \leq \sup_{u \in \Hscr(1)} Q(\bm \truew, \lambda,  u) + \eta R(\bm \truew, \lambda).
        \end{align*}
        Therefore, we have 
        \begin{align*}
       \sup_{u \in \Hscr(1)} Q(\bm \estw, \lambda, u)  \leq \left\{  \sup_{u \in \Hscr(1)} Q(\bm \truew, \lambda,  u) + \eta R(\bm \truew, \lambda) \right\} = \bigOp\left[ (1 + \eta) \frac{\mathcal{N}(\lambda)}{n} + \lambda\right],\\
      R(\bm \estw, \lambda)  \leq \eta^{-1} \left\{ \sup_{u \in \Hscr(1)} Q(\bm \truew, \lambda, u) + \eta R(\bm \truew, \lambda)   \right\} = \bigOp\left[ \frac{\mathcal{N}(\lambda)}{n} + \eta^{-1}\left(\lambda + \frac{\mathcal{N}(\lambda)}{n} \right) \right].
      \end{align*}
      \end{proof}

    \begin{proof}[\bf Proof of Theorem \ref{thm:emp_what_regress}]
      By Theorem \ref{thm:emp_bal_error_truew}, we can derive that 
      \begin{align*}
        Q(\bm \estw, \lambda,  m)  = \bigOp\left[ \|m\|^2_{\Hscr}\left( \frac{\mathcal{N}(\lambda)}{n} + \lambda  + \eta \frac{\mathcal{N}(\lambda)}{n} \right)\right],\\
        R(\bm \estw, \lambda, ) = \bigOp\left[ \eta^{-1}\left(  \frac{\mathcal{N}(\lambda)}{n} + \lambda \right)   +  \frac{\mathcal{N}(\lambda)}{n} \right].
      \end{align*}
    From the decomposition, we have 
     \begin{align}
      \|\hat \tau - \tau\|_n =& \left\| \left( \frac{1}{n}\sum_{i=1}^n \mathcal{S}_{\hat A_i} + \lambda \mathcal{I} \right)^{-1}\left( \frac{1}{n}\sum_{i=1}^n \mathcal{K}_{\hat A_i} \estw_i Y_i\right) - \tau \right\|_n \nonumber\\
      = &\left\|\left( \frac{1}{n}\sum_{i=1}^n \mathcal{S}_{\hat A_i} + \lambda  \mathcal{I}\right)^{-1}\left( \frac{1}{n}\sum_{i=1}^n \mathcal{K}_{\hat A_i} \estw_i m(\hat A_i, \bm X_i)\right) - \frac{1}{n}\sum_{j=1}^n m(\cdot, \bm X_j)\right\|_n \label{eqn:empnorm_decomp1}\\
      & + \left\| \frac{1}{n}\sum_{j=1}^n m(\cdot, \bm X_j) - \tau \right\|_n  \label{eqn:empnorm_decomp2} \\
      & + \left\| \left( \frac{1}{n}\sum_{i=1}^n \mathcal{S}_{\hat A_i} + \lambda  \mathcal{I}\right)^{-1}\left( \frac{1}{n}\sum_{i=1}^n \mathcal{K}_{\hat A_i} \estw_i \epsilon_i\right) \right\|_n \label{eqn:empnorm_decomp3}
     \end{align}
     $$\eqref{eqn:empnorm_decomp1} = \{Q(\bm \estw,  \lambda, m)\}^{1/2} = \bigOp\left[ \|m\|_{\Hscr}\left( \frac{\mathcal{N}(\lambda)}{n} + \lambda  + \eta \frac{\mathcal{N}(\lambda)}{n}\right)^{1/2} \right].$$
     From the bound of \eqref{eqn:comp3} in Theorem \ref{thm:emp_bal_error_truew}, we have 
     $$\eqref{eqn:empnorm_decomp2} = \bigOp\left( \frac{\|m\|_\Hscr}{\sqrt{n}} \right). $$
     Notice that under Assumption \ref{assum:errors}, 
     \begin{align*}
       \E \left\{ \left\| \left( \frac{1}{n}\sum_{i=1}^n \mathcal{S}_{A_i} + \lambda \right)^{-1}\left( \frac{1}{n}\sum_{i=1}^n \mathcal{K}_{A_i} \estw_i \epsilon_i\right) \right\|^2_n \mid A_i, i = 1,\dots,n   \right\}\\
       \leq \sigma_0^2 R(\bm \estw, \lambda) = \sigma_0^2 \bigOp\left[ \eta^{-1} \left( \frac{\mathcal{N}(\lambda)}{n} + \lambda  \right)  +  \frac{\mathcal{N}(\lambda)}{n} \right].
     \end{align*}
    Then $$\eqref{eqn:empnorm_decomp3} = \bigOp\left( \sigma_0 \eta^{-1/2} \left( \frac{\mathcal{N}(\lambda)}{n} + \lambda  \right) ^{1/2} + \left( \frac{\mathcal{N}(\lambda)}{n}  \right)  ^{1/2} \right).$$
  
      Under Assumption \ref{assum:eigen_decay} and the conditions of \(\eta \) and \(\lambda\) stated in Theorem \ref{thm:emp_what_regress}, we note that \(\mathcal{N}(\lambda) \asymp \lambda^{-1/b}\). Follow the proof of Lemma S7 in \cite{wang2020low}, to satisfy the condition \(\sqrt{ \sum_{l=1}^\infty \min \{t_l,\lambda\} /(n\lambda^2)} = \bigO(1)\), we need \(n\lambda^{-(1+1/b)} = \bigO(1)\), which is the same condition for \(\mathcal{N}(\lambda)(\lambda n)^{-1} = \bigO(1)\). Then we can see that when \(\lambda \asymp n^{-b/(1+b)}\), the conditions are satisfied and \(\lambda \asymp (\mathcal{N}(\lambda) /\sqrt{n}\). 
    
   The bound of \(\|\hat \tau - \tau\|_n\) follows.
  
    \end{proof}

\bibliographystyle{chicago}
\bibliography{refer.bib}

\end{document}